\documentclass{aa}
\usepackage{txfonts}
\usepackage{graphicx}
\usepackage{subcaption}
\usepackage{threeparttable} 
\usepackage[colorlinks=true,allcolors=blue]{hyperref}
\usepackage{gensymb}

\DeclareUnicodeCharacter{5415}{}  
\DeclareUnicodeCharacter{5EFA}{}  
\DeclareUnicodeCharacter{4F1F}{}  

\begin{document} 

\authorrunning{Mountrichas et al.}
\titlerunning{AGN environments vs. black hole mass, accretion, and luminosity}

\title{Large-scale environments of star-forming active galactic nuclei: How black hole mass, accretion rate, and luminosity connect to dark matter halos}

\author{G. Mountrichas\inst{1}, F. J. Carrera\inst{1}, F. Shankar\inst{2}, A. Georgakakis\inst{3} }
          
     \institute {Instituto de Fisica de Cantabria (CSIC-Universidad de Cantabria), Avenida de los Castros, 39005 Santander, Spain.
              \email{gmountrichas@gmail.com}
               \and
             Department of Physics and Astronomy, University of Southampton, Highfield, Southampton, SO17 1BJ, UK
              \and
              Institute for Astronomy \& Astrophysics, National Observatory of Athens, V. Paulou \& I. Metaxa, 11532, Greece
             }

\abstract{Understanding the relative roles of large-scale environment and internal host-galaxy processes in shaping AGN activity is key to constraining models of black-hole growth and galaxy evolution. 
In this work, we investigate how the environment of X-ray selected active galactic nuclei (AGN) relates to black-hole growth and accretion properties, and whether these introduce a dependence on large-scale environment beyond that set by the host galaxy itself. 
By combining the XXL and Stripe~82X surveys, we assemble samples of 427 broad-line AGN at $0.5<z<1.2$ and over $20\,000$ galaxies, with host-galaxy properties consistently derived using the same spectral energy distribution fitting methodology and assumptions. 
Dark matter halo (DMH) masses are inferred from AGN--galaxy cross-correlation functions, while a multivariate nearest-neighbour matching algorithm is applied to isolate trends with black-hole mass ($M_{\mathrm{BH}}$), Eddington ratio ($\lambda_{\mathrm{Edd}}$), and X-ray luminosity ($L_{\mathrm{X}}$) under controlled host-galaxy conditions. Within the statistical uncertainties of the present dataset, we find that X-ray AGN typically reside in halos of $\log(M_{\mathrm{DMH}}/h^{-1}M_\odot)\simeq13$, with no significant variation as a function of $M_{\mathrm{BH}}$, $\lambda_{\mathrm{Edd}}$, or $L_{\mathrm{X}}$. 
Neither $M_{\mathrm{BH}}$ nor $\lambda_{\mathrm{Edd}}$ shows a measurable correlation with large-scale environment, consistent with a scenario in which long-term black-hole growth and short-term accretion variability are primarily regulated by internal, host-galaxy processes rather than by halo-scale mass alone. 
The absence of a statistically significant $M_{\mathrm{DMH}}$--$L_{\mathrm{X}}$ trend further indicates that AGN radiative output reflects stochastic or feedback-regulated variability, rather than direct modulation by the large-scale environment. 
Overall, these results support a self-regulated co-evolution framework in which large-scale structure sets the boundary conditions for gas availability and AGN duty cycle, while the subsequent growth and luminosity evolution of AGN are determined predominantly by local processes within their host galaxies.

}

   
\maketitle  

\section{Introduction}
\label{sec_intro}

Active galactic nuclei (AGN) represent one of the most energetic phenomena in the Universe, powered by accretion of matter onto supermassive black holes (SMBHs) residing at the centers of galaxies. Their radiative output spans many orders of magnitude, influencing both the host galaxy and the surrounding intergalactic medium through feedback processes that can regulate star formation and baryon cycling on multiple scales. Understanding the environmental context of AGN is therefore crucial for constraining how black-hole growth connects to galaxy evolution across cosmic time.  

The large-scale environment, traced through the dark matter halo (DMH) mass, provides a statistical measure of how AGN are distributed within the cosmic web and how AGN host galaxies populate large-scale structure.
The clustering strength of AGN reflects the typical DMH mass associated with their host galaxies, and its dependence on host and SMBH properties offers insight into whether observed environmental trends arise from large-scale structure or from internal, secular processes within galaxies. Theoretical and semi-empirical models have also emphasized the role of halo assembly in shaping the long-term growth of supermassive black holes and the global regulation of feedback \citep[e.g.][]{Shankar2010, Shankar2013, Shankar2017, Shankar2020, Hopkins2006, Hopkins2008, Croton2006, Bower2006, Fanidakis2012, Hirschmann2014, Weinberger2018}.  In these frameworks, the mass and assembly history of the DMH primarily influence the availability of gas, the probability of triggering AGN episodes, and the duty cycle of accretion, while the instantaneous accretion rate and radiative output are largely governed by baryonic processes internal to the host galaxy, such as gas inflows, disc instabilities, and feedback self-regulation. even if instantaneous accretion rates are determined by local conditions. Quantifying these dependencies is therefore a key step toward disentangling environmental and host-galaxy effects in the co-evolution of galaxies and SMBHs.

It is important to stress that large-scale ($>1\,\mathrm{Mpc}$) clustering measurements do not directly trace AGN triggering mechanisms. Instead, they primarily reflect the properties of the host galaxy population, most notably stellar mass ($M_\star$), and the way AGN occupy the galaxy population \citep[e.g.][]{Leauthaud2015, Georgakakis2019, Allevato2021, Shankar2020}. In this context, AGN are known to occupy a broad distribution of dark matter halo masses that peaks at a few $\times10^{12}$\,M$_\sun$, while the “typical” halo masses inferred from correlation function and bias measurements are biased toward higher values because they are sensitive to the high-mass tail of this distribution \citep[e.g.,][]{Mountrichas2013}. As a result, differences in the inferred halo masses among AGN samples can naturally arise from selection effects, namely how different observational criteria preferentially select hosts with different M$_\star$, rather than reflecting distinct AGN triggering mechanisms.

Previous studies have explored how AGN clustering depends on a variety of host and black-hole properties, though results often differ due to sample selection, redshift coverage, and the lack of consistent control of covariates.  Regarding host-galaxy properties, a mild dependence of clustering amplitude on M$_\star$ has been reported (\citealt{Mountrichas2019}), while other analyses found no such trend (\citealt{Allevato2019, Viitanen2019}).  
These latter studies, which probed broad redshift intervals up to $z\approx3$ and $z\approx2.5$, suggest that the environment–M$_\star$ connection may weaken at higher redshift or depend on AGN type. In contrast, a dependence on star-formation rate (SFR) has been observed (\citealt{Mountrichas2019, Allevato2019}), with the strongest effect seen for the specific star-formation rate (sSFR), where AGN in less star-forming galaxies occupy more massive halos (\citealt{Mountrichas2019}).

The situation for X-ray luminosity ($L_{\mathrm{X}}$) is more complex.  
A positive correlation between clustering amplitude and $L_{\mathrm{X}}$ has been identified at low redshifts ($z<0.5$; \citealt{Krumpe2010b, Krumpe2012}), whereas no dependence is observed in the local Universe ($0.007<z<0.037$), although differences appear in how low- and high-luminosity AGN populate halos on small scales (\citealt{Krumpe2018}).  
At intermediate redshifts, a mild dependence has been suggested (\citealt{Powell2020}), but other analyses find an inverse relation, where more luminous AGN reside in less massive halos (\citealt{Mountrichas2016}).  
Such discrepancies are often attributed to differences in luminosity intervals, redshift coverage, and selection effects (\citealt{Fanidakis2013a, Fanidakis2013b, Powell2020}).

Dependence on black-hole mass ($M_{\mathrm{BH}}$) appears weak, with only mild trends found up to $z<0.6$ (\citealt{Krumpe2015, Krumpe2023}). Studies of optical quasars showed that when the sample is divided by $M_{\mathrm{BH}}$, clustering differences emerge only for the most massive subset (\citealt{Shen2009}).  
In contrast, the Eddington ratio ($\lambda_{\mathrm{Edd}}$), which traces instantaneous accretion efficiency, exhibits no clear correlation with environment up to comparable redshifts (\citealt{Krumpe2015, Krumpe2023}).  
Using its observational proxy, the specific black-hole accretion rate ($L_{\mathrm{X}}/M_\star$), no significant dependence is detected up to $z\sim2.5$ (\citealt{Viitanen2019}) or $z\sim3$ for type~2 AGN (\citealt{Allevato2019}). 

Obscuration introduces additional ambiguity: some studies report no clustering difference between obscured and unobscured AGN (\citealt{Viitanen2023}), while others find unobscured systems to be more clustered (\citealt{Allevato2014}) or obscured AGN to reside in more massive halos (\citealt{Powell2018}); in the local Universe, type~1 and type~2 AGN may even occupy halos differently (\citealt{Krumpe2018}).

A key limitation of many previous studies is that, when investigating the dependence of clustering on one parameter, other potentially relevant host or SMBH properties are not simultaneously controlled. Some analyses account for one or two covariates (e.g. M$_\star$ or $\lambda_{\mathrm{Edd}}$), but few match across multiple physical dimensions, which can introduce biases and mask genuine correlations.  
Moreover, most existing works are confined to low redshifts ($z<0.6$) or span wide redshift ranges that complicate interpretation due to possible evolutionary mixing.

Within this framework, our goal is not to identify AGN triggering mechanisms, but to quantify how different AGN selections probe the underlying halo-mass distribution once both SMBH and host-galaxy properties are carefully controlled. To this end, we use X-ray selected AGN from the XMM--XXL and Stripe~82X surveys to examine the relation between the typical dark-matter halo mass and $M_{\mathrm{BH}}$, $\lambda_{\mathrm{Edd}}$, and $L_{\mathrm{X}}$.

A multivariate nearest-neighbour matching algorithm is employed to explicitly control for host-galaxy $M_\star$, as well as SFR, sSFR, redshift, and the remaining SMBH parameters not under investigation in each case. This ensures that, when comparing AGN subsets split by a given parameter, their $M_\star$ and other covariate distributions are statistically consistent. As a result, any differences in the inferred halo masses cannot be attributed to systematic variations in host-galaxy properties.

The paper is structured as follows.  
Section~\ref{sec_data} describes the AGN and galaxy samples. Section~\ref{sec_analysis} presents the derivation of physical parameters from spectral energy distribution (SED) fitting, the clustering analysis and matching methodology.  
Results on the dependence of halo mass on $M_{\mathrm{BH}}$, $\lambda_{\mathrm{Edd}}$, and $L_{\mathrm{X}}$ are presented in Section~\ref{sec_results}.  
Section~\ref{sec_discussion} discusses the physical interpretation of these results in the context of AGN and galaxy co-evolution, and Section~\ref{sec_summary} summarizes the main conclusions.

Throughout this paper, we adopt a flat $\Lambda$CDM cosmology with 
$\Omega_{\mathrm{M}}=0.315$, $\Omega_{\Lambda}=0.685$, $h=0.674$ (i.e.\ $H_0 = 67.4~\mathrm{km\,s^{-1}\,Mpc^{-1}}$), and 
$\sigma_8(z=0) = 0.811$
consistent with the \citet{Planck2020} cosmological parameters.

\section{Data}
\label{sec_data}

In this study, we use data from two well-studied extragalactic survey fields, namely XMM--XXL and Stripe 82X. Below, we describe the main characteristics of the X-ray AGN and galaxies observed in these fields.

\subsection{X-ray AGN}
\label{sec_agn}

Our analysis is based on X-ray selected AGN detected in the XMM--XXL North (hereafter XXL) and Stripe 82X fields. The XXL AGN sample has been extensively used and described in several of our previous works \citep[e.g.][]{Masoura2018, Masoura2021, Pouliasis2020, Mountrichas2023a, Mountrichas2023b, Mountrichas2023d, Mountrichas2024d}. 

In brief, XMM--XXL \citep[][]{Pierre2016} is a medium-depth X-ray survey covering a total area of $\sim$50\,deg$^2$, divided into two nearly equal regions: the XMM--XXL North and South fields. The XXL-N dataset includes 8445 X-ray sources, of which 5294 have optical counterparts in the SDSS and 2512 have reliable spectroscopic redshifts \citep[][]{Menzel2016, Liu2016}. Mid-infrared (MIR) and near-infrared (NIR) counterparts were identified using the likelihood ratio method \citep[][]{Sutherland_and_Saunders1992}, as implemented in \citet{Georgakakis2011}. Details on the XMM-Newton data reduction and infrared counterpart identification are provided in \citet{Georgakakis2017}.

Among the 2512 XXL-N X-ray sources with reliable spectroscopy, 1786 have been classified as broad-line AGN (BLAGN1) by \citet[][]{Menzel2016}. A source was identified as BLAGN1 when at least one broad emission line had a full width at half-maximum (FWHM) greater than 1000\,km\,s$^{-1}$. \citet[][]{Liu2016} performed spectral fitting of the BOSS spectroscopy for these 1786 BLAGN1 to estimate single-epoch virial black hole masses ($M_{\mathrm{BH}}$) from the continuum luminosities and broad-line widths, following the approach of \citet[][]{Shen2013}. 

In short, they measured the continuum luminosities and the FWHMs of the broad emission lines, and then applied the appropriate single-epoch virial mass estimators depending on redshift: H$\beta$ for $z<0.9$, Mg\,\textsc{ii} for $0.9<z<2.2$, and C\,\textsc{iv} for $z>2.2$. Previous studies have demonstrated that $M_{\mathrm{BH}}$ derived from different broad lines using these fiducial relations are generally consistent with each other, with negligible systematic offsets and small scatter \citep[e.g.][]{Shen2008, Shen2011, Shen2013, Shen2012}. \citet[][]{Liu2016} confirmed this consistency for the XXL sample. The average uncertainty of their $M_{\mathrm{BH}}$ measurements is $\sim$0.5\,dex, while sources with higher signal-to-noise ratio spectra have errors below 0.15\,dex.

The Stripe 82X survey \citep[][]{LaMassa2013a, LaMassa2013b, LaMassa2015} was designed to exploit the extensive multiwavelength coverage of Stripe 82, with the main goal of identifying high-luminosity AGN at high redshift. The survey includes both \textit{Chandra} and XMM--Newton observations. In this work, we use the Data Release~3 (DR3) of Stripe 82X, presented by \citet{LaMassa2024}. This latest release combines the X-ray detections from the initial Stripe 82X catalog \citep[DR1;][]{LaMassa2016b} with the multiwavelength identifications and photometric redshifts from \citet[DR2;][]{Ananna2017}, as well as spectroscopic redshifts from independent surveys, the dedicated SDSS-IV eBOSS program \citep[][]{LaMassa2019}, and additional ground-based observations obtained with the Palomar and Keck telescopes. 

The Stripe 82X DR3 catalog covers a total area of 31.3\,deg$^2$ and contains 6181 X-ray sources. Spectroscopic redshifts are available for 3457 of them, corresponding to an overall spectroscopic completeness of 56\%. Within the contiguous regions of homogeneous X-ray coverage and for sources brighter than $r=22$\,mag, the spectroscopic completeness increases to 90\%. The catalog adds 343 new spectroscopic redshifts compared to DR2 and includes $M_{\mathrm{BH}}$ estimates derived from SDSS spectra of 1297 Type~1 AGN, which we use in our analysis and describe below.

Compared to earlier releases, DR3 introduces additional columns that facilitate AGN identification based on X-ray luminosity. Sources are classified as X-ray AGN when their rest-frame 2--10\,keV luminosity exceeds $10^{42}$\,erg\,s$^{-1}$ \citep[][]{BrandtHasinger2005}. The rest-frame luminosities are $k$-corrected using a power-law spectral model with slopes $\Gamma=2.0$ (soft band) and $\Gamma=1.7$ (hard and full bands), following the procedure outlined by \citet[][]{LaMassa2013b, LaMassa2016b}. Detection thresholds correspond to significance levels of $4.5\sigma$ for \textit{Chandra} and $5\sigma$ for \textit{XMM-Newton} sources. When a source is not detected in the hard band, its 2--10\,keV luminosity is estimated by scaling the full- or soft-band luminosities, as detailed in \citet[][]{LaMassa2019}. 

Finally, we note that in our analysis, for both X-ray AGN datasets, we use the observed, i.e., not corrected for absorption, $2-10$\,keV $L_X$ provided in the two catalogues.

\subsection{Galaxy samples}
\label{sec_galaxies}

To investigate the large-scale environments of X-ray AGN, we estimate the mass of their host DMHs through the two-point cross-correlation function. For this purpose, we cross-correlate the AGN samples with galaxy catalogues that spatially overlap with the X-ray survey regions. 

For the XXL AGN, we follow the methodology described in \citet{Mountrichas2016, Mountrichas2019} and use the spectroscopic galaxy catalogue from the VIMOS Public Extragalactic Redshift Survey \citep[VIPERS;][]{Garilli2014, Guzzo2014}. This dataset has also been employed in several of our previous (non clustering) studies \citep[e.g.][]{Pouliasis2020, Mountrichas2023d, Mountrichas2024d}.

In summary, the galaxy sample used in our analysis is drawn from the second public data release (PDR-2; \citealt{Scodeggio2018}) of VIPERS. The survey was carried out with the Visible Multi-Object Spectrograph \citep[VIMOS;][]{LeFevre2003} on the ESO Very Large Telescope (VLT) and covers an effective area of $\approx$23.5\,deg$^2$, split between two regions (W1 and W4) within the Canada–France–Hawaii Telescope Legacy Survey–Wide (CFHTLS-Wide) fields. Spectroscopic targets were selected down to a magnitude limit of $i_{\mathrm{AB}}<22.5$ using the T0006 release of the CFHTLS photometric catalogues. An optical colour–colour preselection, defined as $(r-i) > 0.5(u-g)$ or $(r-i) > 0.7$, was applied to remove low-redshift ($z<0.5$) galaxies, achieving a completeness greater than 98\% for $z>0.5$ \citep[see][]{Guzzo2014}. 

The PDR-2 catalogue contains 86\,775 galaxies with spectroscopic redshifts. Each spectrum is assigned a quality flag that quantifies the reliability of the redshift measurement. Following the standard VIPERS convention, only galaxies with flags between 2 and 9 are considered to have reliable redshifts and are included in our analysis \citep[][]{Garilli2014, Scodeggio2018}. Restricting the sample to the redshift range $0.5<z<1.2$ yields a final set of 45\,180 galaxies.

For the Stripe 82X field, we make use of galaxies located within the Stripe 82 region on the celestial equator, which spans approximately 300\,deg$^2$, with right ascension ranging from 20$^{\mathrm{h}}$ to 04$^{\mathrm{h}}$ and a declination of $\pm1.25^{\circ}$. Stripe 82 is part of the Sloan Digital Sky Survey \citep[SDSS;][]{Jiang2014} and benefits from deep, repeated imaging that has been co-added to produce significantly deeper data than single-epoch SDSS observations. The co-added images reach 2–3\,mag deeper, depending on the photometric band, allowing the detection of faint galaxies and studies of the low surface brightness Universe \citep[][]{Annis2014}. 

In addition to the SDSS optical data, the Stripe 82 region has extensive multiwavelength coverage, including near-infrared (NIR) imaging from the UKIRT Infrared Deep Sky Survey \citep[UKIDSS;][]{Lawrence2007}, mid-infrared (MIR) observations from the \textit{Spitzer Space Telescope}, and far-infrared coverage from \textit{Herschel} \citep[][]{Viero2014}. Galaxy redshifts are primarily spectroscopic, supplemented where necessary by high-quality photometric redshifts derived from the extensive multiwavelength coverage available in this field. For the purpose of our clustering analysis, we restrict the galaxy sample to the regions of Stripe 82 that overlap with the X-ray footprint of the Stripe 82X survey (see Section~\ref{sec_selection_criteria}) and to the same redshift range used for the AGN–galaxy cross-correlation measurements, ensuring consistency with the XXL analysis. The resulting redshift distribution closely matches that of the XXL/VIPERS galaxy sample, with similar median redshifts (0.70 for Stripe 82 and 0.65 for XXL/VIPERS), ensuring that systematic differences in redshift sampling do not bias the inferred halo masses.

\section{Analysis}
\label{sec_analysis}

In this section, we present the SED fitting procedure and the clustering analysis used to derive the host-galaxy properties and the DMH masses of our samples. We also describe the selection criteria applied to define the final X-ray AGN and galaxy datasets.

\subsection{SED fitting}
\label{sec_sed_fitting}

To derive the host galaxy properties and estimate the $\lambda_{\mathrm{Edd}}$ of the AGN, we construct and fit their SEDs. To ensure reliable results, we follow the methodology established in our previous studies \citep[e.g.][]{Mountrichas2021b, Mountrichas2021c, Mountrichas2022a, Mountrichas2022b, Mountrichas2022c, Mountrichas2023c, Mountrichas2024a, Mountrichas2024b}, requiring all AGN and galaxy sources to have detections in the optical, near-infrared (NIR), and mid-infrared (MIR) photometric bands.

The SED construction and fitting for most datasets have been presented in our previous works. Specifically, the SEDs of the XXL AGN were built, fitted, and analysed in \citet[][]{Mountrichas2023a, Mountrichas2023b, Mountrichas2023d, Mountrichas2024d}, while those of the VIPERS galaxies were described in \citet[][]{Mountrichas2023d, Mountrichas2024d}. The SEDs of galaxies in the Stripe82 region were modelled in \citet[][]{Mountrichas2025a}. Further details on the photometric coverage, parameter grids, and fitting configuration can be found in these references. In the present work, we perform SED fitting only for the X-ray AGN detected in Stripe 82X.

For the SED fitting, we adopt the same set of templates and parameter grid as in our previous analyses. In summary, the galaxy component is described by a delayed star formation history (SFH) of the form $\mathrm{SFR}(t) \propto t \times \exp(-t/\tau)$. A secondary burst component is included as a constant star formation episode lasting 50\,Myr, following the prescription of \citet[][]{Malek2018} and \citet[][]{Buat2019}. Stellar emission is modelled using the single stellar population (SSP) templates of \citet[][]{Bruzual_Charlot2003} and is attenuated according to the \citet[][]{Charlot_Fall_2000} law. Nebular emission is modelled using templates based on \citet[][]{VillaVelez2021}, while the dust emission heated by stars is represented by the \citet[][]{Dale2014} models, excluding any AGN contribution. The AGN emission is included using the \texttt{SKIRTOR} models \citep[][]{Stalevski2012, Stalevski2016}, which account for both the torus and the central engine components. The complete list of modules and the adopted parameter space used for the SED fitting can be found, for example, in Table~1 of \citet[][]{Mountrichas2022a, Mountrichas2022c}.

\subsection{Selection criteria and final datasets}
\label{sec_selection_criteria}

\begin{figure}
\centering
  \includegraphics[width=0.8\columnwidth, height=5.5cm]{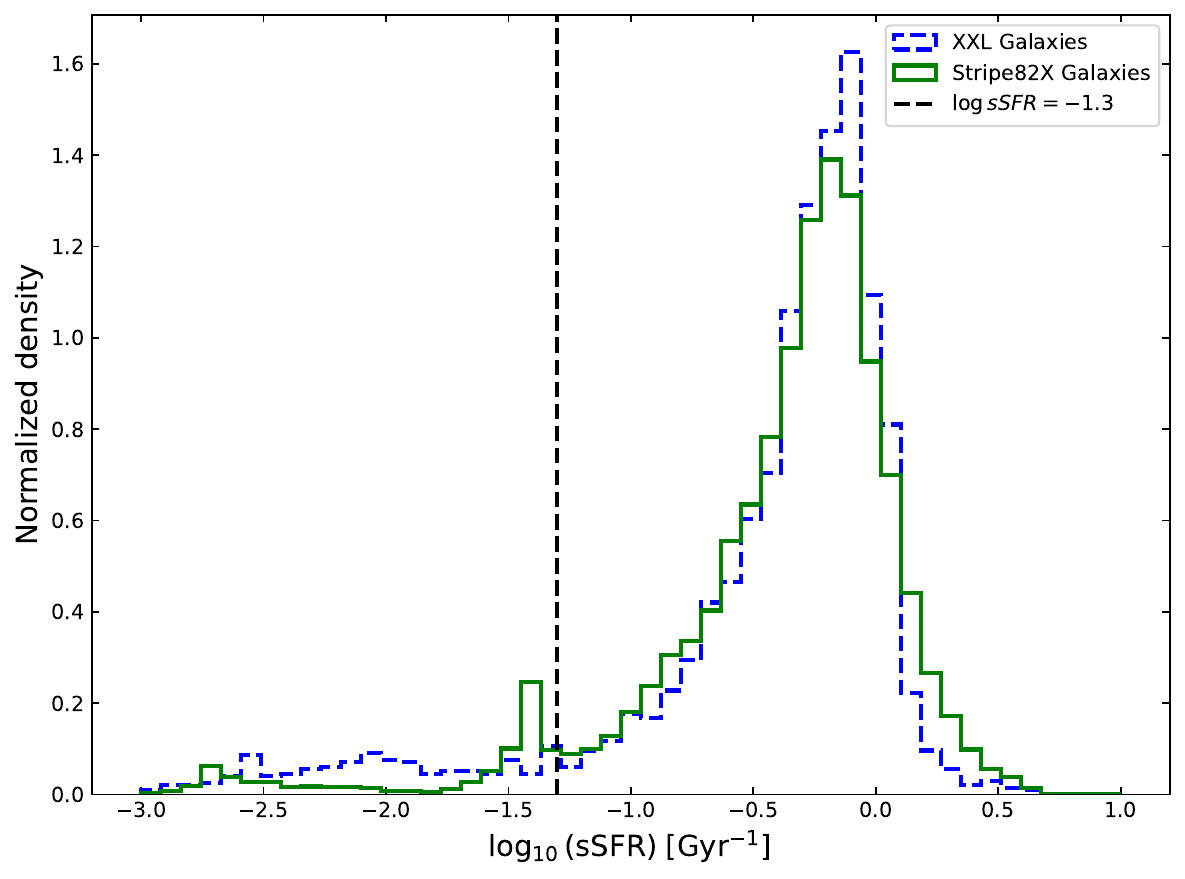}  
  \includegraphics[width=0.8\columnwidth, height=5.5cm]{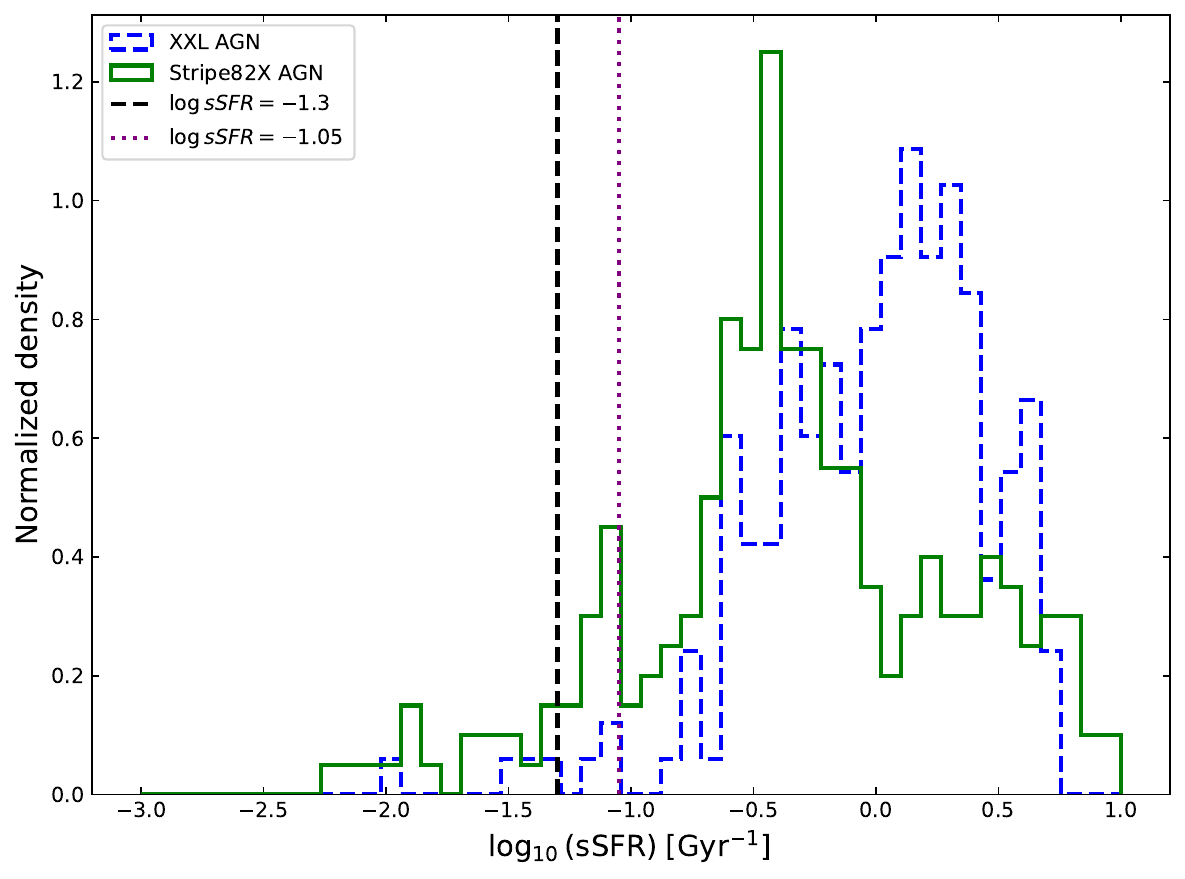}  
  \caption{Distributions of the sSFR ($\mathrm{sSFR} = \mathrm{SFR}/M_\star$), for galaxies (top panel) and X-ray AGN (bottom panel) for the two fields, as indicated in the legend.}
  \label{fig_ssfr}
\end{figure} 

As discussed in the previous section, we require all datasets to have photometric coverage in the optical, NIR, and MIR bands. This criterion reduces the number of VIPERS galaxies to 14\,128 sources \citep[see Sect.~2.2 in][]{Mountrichas2023d}, while 320\,319 galaxies in Stripe82 fulfil the same requirement \citep[see Sect.~2.2 in][]{Mountrichas2025a}. Among the X-ray samples, 687 XXL AGN and 731 Stripe 82X AGN meet this photometric criterion and have available $M_{\mathrm{BH}}$ measurements. 

Following our previous studies \citep[e.g.][]{Mountrichas2021c, Buat2021, Koutoulidis2022, Pouliasis2020}, we excluded sources with poor SED fits or unreliable host galaxy estimates. Specifically, we applied a reduced chi-square threshold of $\chi^{2}_{\mathrm{r}} \leq 5$ \citep[e.g.][]{Masoura2018, Buat2021} and removed systems for which \texttt{CIGALE} could not constrain key physical parameters such as the SFR and $M_\star$. To identify these cases, we adopted the same reliability criteria used in our recent analyses \citep[e.g.][]{Mountrichas2021b, Mountrichas2022b, Mountrichas2022c, Buat2021, Koutoulidis2022, Pouliasis2022}. For each parameter estimated by \texttt{CIGALE}, two values are provided: one corresponding to the best-fitting model and another representing the Bayesian likelihood-weighted mean. Large discrepancies between these values indicate a complex likelihood distribution and significant uncertainties. We therefore retained only sources satisfying the conditions $1/5 \leq \mathrm{SFR}_{\mathrm{best}} / \mathrm{SFR}_{\mathrm{bayes}} \leq 5$ and $1/5 \leq M_{\ast,\mathrm{best}} / M_{\ast,\mathrm{bayes}} \leq 5$, where the subscripts “best” and “bayes” denote the best-fit and Bayesian estimates, respectively. After applying these criteria, the remaining sample consists of 181\,361 galaxies in Stripe82 and 10\,150 in VIPERS. For the X-ray samples, 584 AGN in XXL and 550 in Stripe 82X satisfy these quality constraints. 

Furthermore, to ensure that the galaxy catalogues represent non-active systems, we excluded potential AGN candidates, i.e., galaxies that are not detected in X-rays but exhibit a strong AGN component in their SEDs. We removed all galaxies with an AGN fractional contribution $\mathrm{frac_{AGN}} > 0.2$ \citep[e.g.][]{Mountrichas2021c, Mountrichas2022a, Mountrichas2022b, Mountrichas2022c, Mountrichas2024d}, where $\mathrm{frac_{AGN}}$ denotes the fraction of the total infrared emission attributed to AGN heating. This criterion excludes approximately 45\% of VIPERS galaxies and 35\% of Stripe 82 galaxies, consistent with the fractions reported in our previous analyses of the Boötes, COSMOS, and eFEDS fields, at similar redshifts \citep[][]{Mountrichas2021c, Mountrichas2022a, Mountrichas2022b}. 

In addition, we restrict all samples to the redshift range
$0.5 \leq z \leq 1.2$, consistent with the redshift coverage of VIPERS (see Sect.~\ref{sec_galaxies}). The final datasets comprise $\sim$4,000 galaxies in VIPERS and $\sim$40,000 in Stripe 82, along with 240 X-ray AGN in XXL and 290 in Stripe 82X within the same redshift interval. The much larger galaxy sample in Stripe 82 arises from the substantially deeper optical imaging, which reaches $\sim$2–2.5\,mag fainter than the CFHTLS–W1 photometry used in the XXL area. This allows Stripe 82 to detect a far larger number of low-mass and moderately star-forming galaxies at $z\sim1$,
whereas X-ray AGN counts are governed primarily by the depth of the X-ray surveys. Since the XMM coverage in XXL and Stripe 82X is similar in sensitivity and sky area, the number of detected AGN is comparable across the two fields despite the large difference in galaxy counts.

To prepare for the clustering analysis, we ensured that AGN and galaxies occupy the same effective survey regions within each field by matching their spatial footprints using a custom \texttt{Python} routine. 
For each AGN, we identify galaxies located within a projected comoving separation $r_p \leq 20\,h^{-1}\,\mathrm{Mpc}$ and a line-of-sight separation $\pi \leq 60\,h^{-1}\,\mathrm{Mpc}$. 
This three-dimensional window is chosen to be sufficiently large to guarantee that, for all transverse scales used in the $w_p(r_p)$ measurements ($r_p \leq 30\,h^{-1}\,\mathrm{Mpc}$), galaxy neighbours can be measured around each AGN without being affected by survey edges or masked regions.

AGN that have no galaxies within this broad geometric window are removed from the analysis. Such cases indicate that the AGN lies outside the effective galaxy footprint (e.g. in masked or sparsely sampled areas) rather than reflecting a physically underdense environment. 
Importantly, no minimum number of galaxies is required: the criterion is purely geometric and does not depend on the local galaxy density in a physical sense. The statistical robustness of the clustering signal is handled at the level of pair counts, fitting ranges, and jackknife error estimation.

After applying this footprint-matching requirement, the samples used in the clustering analysis consist of 203 and 245 X-ray AGN, and 2\,428 and 19\,141 galaxies in XXL and Stripe 82X, respectively. These matched samples probe identical sky regions and redshift intervals, ensuring a consistent and unbiased basis for the AGN–galaxy cross-correlation measurements.

Finally, we restrict our analysis to AGN hosted by star-forming galaxies. Quiescent galaxies are known to reside in more massive halos and have higher clustering amplitudes than star-forming systems at the same $M_\star$ \citep[e.g.][]{Coil2017}, which would introduce an environmental bias unrelated to black-hole mass.
By selecting only star-forming hosts, all AGN subsamples are drawn from a homogeneous population,  allowing us to isolate environmental trends linked specifically to BH properties.

To identify quiescent systems, we examined the distributions of specific star-formation rate (sSFR), defined as $\mathrm{sSFR} = \mathrm{SFR}/M_\star$ (Fig.~\ref{fig_ssfr}). The sSFR distributions of the galaxy samples exhibit a prominent peak at high sSFR values and a secondary peak at lower values, corresponding to quiescent systems \citep[e.g.][]{Mountrichas2021c, Mountrichas2022c}. Using the galaxy samples, owing to their larger statistics, we identify the position of this secondary peak at $\log(\mathrm{sSFR/Gyr^{-1}}) \approx -1.3$ (top panel of Fig.~\ref{fig_ssfr}). We note that the AGN samples display a similar behaviour, with the secondary peak appearing at comparable sSFR values (bottom panel of Fig.~\ref{fig_ssfr}).

However, the sSFR distributions of the two AGN samples differ, with XXL AGN appearing, on average, more star-forming than their Stripe~82X counterparts. This difference can arise even though the two surveys have comparable X-ray flux limits, because the effective selection of the final AGN samples also depends on the identification of optical/IR counterparts and on the availability and quality of redshifts and multiwavelength photometry used for SED fitting, which differ between fields. In particular, sSFR estimates are sensitive to the photometric coverage and depth (e.g. UV/optical constraints on recent star formation and IR constraints on obscured star formation), which are not identical across XXL and Stripe~82. In addition, modest differences in the underlying redshift and stellar-mass distributions, as well as field-to-field variance, can shift the observed sSFR distributions.

Importantly, these differences do not bias our main conclusions. When testing for a dependence on a given AGN property, we apply multivariate matching between the AGN subsamples themselves (Sect.~\ref{sec_matching}), ensuring that their redshift and host-galaxy properties (including 
$M_\star$, SFR, and sSFR) are statistically consistent before comparing their clustering measurements.

\begin{figure*}
\centering
  \includegraphics[height=9.cm]{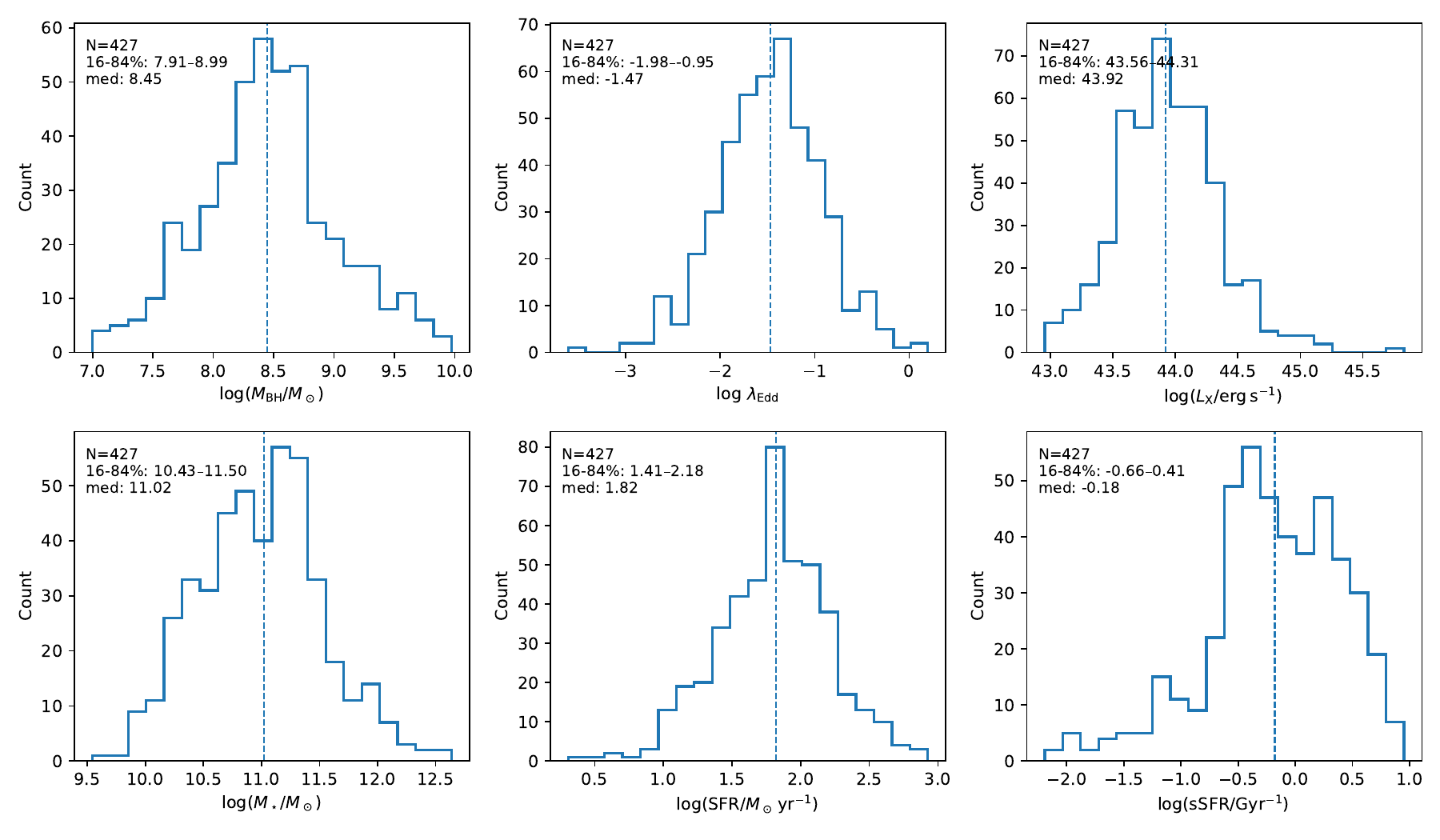}  
  \caption{Distribution of key physical properties for the combined X-ray AGN sample from the XXL and Stripe~82X surveys. From left to right and top to bottom, the panels show the distributions of black-hole mass ($\log M_{\mathrm{BH}}/M_\odot$), Eddington ratio ($\log \lambda_{\mathrm{Edd}}$), X-ray luminosity ($\log L_{\mathrm{X}}/\mathrm{erg\,s^{-1}}$), stellar mass ($\log M_\star/M_\odot$), star-formation rate ($\log \mathrm{SFR}/M_\odot\,\mathrm{yr^{-1}}$), and specific star-formation rate ($\log \mathrm{sSFR}/\mathrm{Gyr^{-1}}$). All quantities are shown in logarithmic units.}
  \label{fig_agn_distrib}
\end{figure*}

Based on the sSFR distributions, we classify as quiescent all systems with $\log(\mathrm{sSFR/Gyr^{-1}}) < -1.3$. Approximately 10\% of the galaxies and 5\% of the AGN in both fields fall below this limit. However, since galaxies serve solely as the tracer population in our cross-correlation analysis, we exclude quiescent systems only from the AGN subsets. Removing them from the galaxy samples would reduce their number density and unnecessarily increase the statistical uncertainties of the clustering measurements. The final number of sources included in the clustering analysis is listed in Table~\ref{table_numbers}. The distributions of black-hole and host-galaxy properties are shown in Fig.~\ref{fig_agn_distrib}.

\begin{table}[ht]
\centering
\caption{Final numbers of sources used in the clustering analysis after applying all photometric, quality, and redshift selection criteria, restricting to overlapping AGN--galaxy regions in each field and excluding quiescent systems from the X-ray datasets.}
\label{table_numbers}
\begin{tabular}{lcc}
\hline\hline
Field & X-ray AGN & Galaxies \\
\hline
XMM--XXL & 199 & 2\,428 \\
Stripe 82X & 228 & 19\,141 \\
\hline
\end{tabular}
\end{table}

\subsection{Clustering analysis}
\label{sec_clustering}

To quantify the large-scale environments of AGN, we measured the AGN--galaxy cross-correlation function, $w_p(r_p)$ \citep[e.g.,][]{Mountrichas2009a, Mountrichas2009b, Hickox2009, Donoso2010,  Krumpe2010a, Miyaji2011, Mountrichas2012, Mountrichas2013, Shen2013, Georgakakis2014, Krumpe2012, Mendez2016, Shirasaki2016, Krumpe2018, Georgakakis2019}. The use of the cross-correlation, instead of the AGN autocorrelation, offers two main advantages. First, it significantly reduces statistical uncertainties, as the galaxy sample provides a denser tracer of the underlying matter distribution. Second, it avoids the need to construct random catalogues for the AGN, whose selection function is typically complex due to spatially varying X-ray sensitivity and survey geometry. Random catalogues are required only for the galaxy populations, whose angular and redshift selection functions are well defined. This approach follows the methodology of \citet{Mountrichas2016, Mountrichas2019}, who applied the same technique to XXL AGN and VIPERS galaxies.

The AGN–galaxy cross-correlation approach assumes that any residual angular incompleteness of the AGN sample is uncorrelated with the large-scale density field traced by galaxies. On the scales considered in this work ($r_p \gtrsim 2$--$3\,h^{-1}\,\mathrm{Mpc}$), this assumption is well justified, as AGN-specific selection effects (e.g. variability, compactness, or optical magnitude limits) are not expected to correlate with large-scale structure. This methodology has been extensively adopted in previous X-ray AGN clustering studies and provides robust estimates of the linear bias and typical dark matter halo mass. 

As an additional validation of the angular selection underlying the AGN–galaxy cross-correlation approach, we performed a test in which the AGN angular correlation function is compared to that of random subsamples of the galaxy catalogue using the same survey mask. The results, presented in Appendix \ref{sec_appendix}, show that the AGN angular clustering is fully consistent with expectations from random galaxy subsamples on the large scales relevant for our bias measurements.

The comoving separation between two objects can be decomposed into components perpendicular ($r_p$) and parallel ($\pi$) to the line of sight. The two-dimensional redshift-space correlation function, $\xi(r_p,\pi)$, was estimated using the \citet{Davis1983} estimator:
\begin{equation}
\xi(r_p, \pi) = \frac{DD(r_p, \pi)}{DR(r_p, \pi)} - 1,
\end{equation}
where $DD$ and $DR$ represent the data–data and data–random pair counts, respectively. For the cross-correlation, $DR$ refers to AGN–random pairs, while for the galaxy autocorrelation it refers to galaxy–random pairs. This estimator requires a random catalogue only for the galaxy sample, which accounts for its angular and redshift selection function.

We note that alternative estimators, most commonly the Landy--Szalay \citep{LZ1993} form, can in principle provide lower-variance measurements and improved edge correction when accurate random catalogues are available for all samples entering the pair counts. In our case, the dominant practical limitation is the construction of an AGN random catalogue that faithfully captures the spatially varying X-ray sensitivity and complex survey geometry. We therefore adopt the Davis--Peebles estimator. Moreover, our bias measurements are derived by fitting the two-halo regime on scales $r_p \gtrsim 2$--$3\,h^{-1}\,\mathrm{Mpc}$, where the correlation function is well described by linear biasing and any residual estimator-dependent effects are sub-dominant for the determination of the large-scale bias and inferred halo mass, which are the quantities of interest here.

The projected correlation function, $w_p(r_p)$, was obtained by integrating $\xi(r_p,\pi)$ along the line of sight:
\begin{equation}
w_p(r_p) = 2 \int_0^{\pi_{\mathrm{max}}} \xi(r_p, \pi) \, d\pi.
\end{equation}

The integration limit $\pi_{\mathrm{max}}$ was determined empirically by varying its value and examining the convergence of $w_p(r_p)$. We adopted $\pi_{\mathrm{max}}=40\,h^{-1}\,\mathrm{Mpc}$ for the XXL field and $\pi_{\mathrm{max}}=20\,h^{-1}\,\mathrm{Mpc}$ for the Stripe 82X field, consistent with the saturation scales observed in our tests (see \citealt{Mountrichas2019}). For the galaxy autocorrelation functions, the adopted $\pi_{\mathrm{max}}$ values are $50$ and $40\,h^{-1}\,\mathrm{Mpc}$ for XXL and Stripe 82, respectively.

Uncertainties in the projected correlation functions were estimated using the jackknife resampling technique (e.g. \citealt{Ross2008}). The survey area was divided into $N_{\mathrm{JK}}=32$ and $48$ spatial subregions for XXL and Stripe 82, respectively. The choice of the number of jackknife regions represents a compromise between 
two requirements. First, each subregion must be sufficiently large compared to 
the maximum projected separation used in the clustering fits, so that the 
jackknife samples capture the variance on the scales of interest. 
Second, the number of subregions must be large enough to allow a stable 
estimation of the covariance matrix. We verified that varying the number of 
jackknife regions within reasonable limits does not significantly affect the 
derived bias values within the statistical uncertainties. For each resampling, one subregion was excluded and the correlation function recalculated, allowing the construction of the covariance matrix
\begin{equation}
C_{ij} = \frac{N_{\mathrm{JK}} - 1}{N_{\mathrm{JK}}} \sum_{k=1}^{N_{\mathrm{JK}}} 
\left[w_p^k(r_{p,i}) - \overline{w_p}(r_{p,i})\right]
\left[w_p^k(r_{p,j}) - \overline{w_p}(r_{p,j})\right].
\end{equation}
The resulting covariance matrix captures both the uncertainties and the bin-to-bin correlations in $w_p(r_p)$ and is used when fitting the large-scale (two-halo) component of the correlation function and converting the best-fitting clustering amplitude into a halo mass.

To derive the AGN bias and the corresponding DMH mass, we fitted the measured projected correlation functions to the theoretical dark matter correlation function, $w_p^{\mathrm{DM}}(r_p)$, computed from the linear matter power spectrum. Following the two-halo approximation, the observed correlation function at large scales can be written as
\begin{equation}
w_p(r_p) = b^2 \, w_p^{\mathrm{DM}}(r_p),
\end{equation}
where $b$ is the linear bias parameter. The galaxy auto-correlation function (ACF) and the AGN--galaxy 
cross-correlation function (CCF) are fitted independently using their 
corresponding jackknife covariance matrices. The galaxy bias $b_g$ is derived 
from the ACF, while the AGN--galaxy bias $b_{Ag}$ is derived from the CCF. 
The AGN bias is then calculated as

\begin{equation}
b_{\rm AGN} = \frac{b_{Ag}}{b_g}.
\end{equation}

In principle, the uncertainties of the ACF and CCF are correlated because the 
same galaxy sample as well as the same underlying large structure are shared in both measurements. However, the galaxy sample is 
significantly larger than the AGN sample and therefore the uncertainty on 
$b_g$ is much smaller than that on $b_{Ag}$. Consequently, the contribution of the covariance between the two measurements to the final uncertainty of $b_{\rm AGN}$ is negligible, and the errors can be safely propagated assuming independent estimates. We verified this explicitly: the fractional uncertainty on the galaxy bias is 
typically $\Delta b_g/b_g \sim 0.08$, compared to 
$\Delta b_{Ag}/b_{Ag} \sim 0.30$. 
Therefore, the uncertainty on $b_{\rm AGN}$ is dominated by the error on 
$b_{Ag}$, and the covariance term between the ACF and CCF has a negligible 
impact on $\Delta b_{\rm AGN}$.

The fits were performed over the range $r_p = 2$–$30\,h^{-1}\,\mathrm{Mpc}$, where the signal is dominated by the two-halo term. The linear bias inferred from Eq.~(5) was then converted into a characteristic dark matter halo mass using the ellipsoidal collapse model of \citet{Sheth2001}, adopting the analytical bias--halo-mass relations of \citet{Bosch2002}. We fit only the two-halo term of the correlation function, as the available AGN samples do not contain sufficient numbers of pairs to robustly constrain the one-halo term. Consequently, our analysis focuses on scales larger than $\sim$2--3\,$h^{-1}$\,Mpc, where linear biasing provides a valid description of the clustering signal.

Finally, the random catalogues were generated for the galaxy samples, following their exact angular footprint and redshift distribution. Each random catalogue contains 20 times more objects than the corresponding galaxy dataset, ensuring smooth pair statistics. We verified that increasing the size of the random catalogues to 30–50 times the galaxy sample does not lead to a measurable reduction in the statistical uncertainties of the projected correlation functions. This is expected, as the error budget is dominated by cosmic variance and the finite size of the AGN samples, rather than by Poisson noise in the random catalogues. We therefore adopt 20$\times$ random catalogues as a conservative and computationally efficient choice.

\subsection{Matched subsamples and parameter control}
\label{sec_matching}

We investigate the dependence of the AGN large-scale environment on three fundamental physical parameters: $M_{\mathrm{BH}}$, $\lambda_{\mathrm{Edd}}$, and $L_{\mathrm{X}}$. For each parameter, the AGN sample is divided into ``low'' and ``high'' subsets, and the corresponding DMH masses are estimated.

A key aspect of this analysis is to control for other host-galaxy and AGN properties known to correlate with clustering amplitude. Previous studies have shown that X-ray AGN clustering depends on quantities such as $L_{\mathrm{X}}$, $M_\star$, SFR, and sSFR \citep[e.g.][]{Krumpe2010a, Georgakakis2014, Mountrichas2016, Mountrichas2019, Allevato2019, Krumpe2023}. Although the detailed form of these dependencies remains debated \citep[e.g.][]{Krumpe2018, Viitanen2019, Powell2020}, we explicitly account for them when isolating the impact of SMBH-related parameters on the large-scale environment.

Moreover, $M_{\mathrm{BH}}$, $\lambda_{\mathrm{Edd}}$, and $L_{\mathrm{X}}$ are themselves mutually correlated \citep[e.g.][]{Krumpe2015, Mountrichas2023d}. To avoid conflating these interdependencies, we explicitly match the distributions of all parameters not under investigation in each test. 
For example, when examining clustering as a function of $M_{\mathrm{BH}}$, we require the high- and low-$M_{\mathrm{BH}}$ subsamples to have statistically consistent distributions in $\lambda_{\mathrm{Edd}}$, $L_{\mathrm{X}}$, redshift, and host-galaxy properties.

In addition, we distinguish between two broad luminosity regimes, defining low-to-moderate luminosity AGN as systems with $\log(L_{\mathrm{X}}/\mathrm{erg\,s^{-1}}) < 44$ and high-luminosity AGN as those with $\log(L_{\mathrm{X}}/\mathrm{erg\,s^{-1}}) \geq 44$. 
This threshold approximately separates moderate-luminosity Seyfert-like AGN from quasar-level activity and is commonly adopted in X-ray AGN clustering studies. 
The luminosity split allows us to test whether any residual environmental trends depend on accretion regime while maintaining sufficient statistics within each subsample.

To construct subsamples with consistent distributions across all control 
parameters, we apply a multivariate nearest-neighbour matching algorithm 
implemented in Python. The matching is performed using a custom routine 
(\texttt{multivar\_match}) developed for this work. The algorithm performs 
a greedy nearest–neighbour search in a multidimensional parameter space 
defined by the matching variables (e.g. $\log(\mathrm{sSFR})$, 
$\log(\mathrm{SFR})$, $\log(M_\star)$, $\log(L_{\mathrm{X}})$, 
$\log(M_{\mathrm{BH}})$, and $\log\lambda_{\mathrm{Edd}}$).

The implementation relies on the NumPy numerical library (\texttt{numpy}) 
and operates on Astropy tables. All covariates are standardized using the 
pooled mean and standard deviation of the two samples, and distances are 
computed using the Euclidean metric in this standardized parameter space. 
For each object in the reference sample, the algorithm identifies the 
closest counterpart in the comparison sample and performs greedy 
nearest–neighbour matching without replacement, ensuring strict 
one-to-one matching between objects. The matching routine is available from the authors upon request.

A key parameter of the method is the caliper, which defines the maximum 
allowable distance between two matched objects in the normalized covariate 
space. We adopt a caliper value of 1.2, which provides a good balance 
between sample size and matching precision. Lower values lead to a 
substantial reduction in the number of matched sources, while higher values 
increase the dispersion of the matched properties and weaken the control 
over covariates. In practice, this procedure ensures that when comparing 
``low'' and ``high'' subsets of a given parameter (e.g. $M_{\mathrm{BH}}$), 
that parameter is allowed to differ by construction, while all other 
covariates remain closely matched within the adopted caliper distance in 
the multivariate parameter space.

For validation, we also experimented with alternative approaches, such as applying hard cuts on the control parameters and matching the distributions of, for instance, $M_\star$, SFR, and sSFR between subsamples. However, these methods resulted in significantly smaller subsets, limiting their statistical usefulness. The multivariate matching approach therefore offers the most efficient compromise between parameter control and sample size. The resulting matched subsamples are consistent in redshift, host-galaxy and SMBH properties, allowing a robust and controlled investigation of how $M_{\mathrm{BH}}$, $\lambda_{\mathrm{Edd}}$, and $L_{\mathrm{X}}$ independently influence AGN clustering.

\section{Results}
\label{sec_results}

This section presents the derived DMH masses of the full X-ray AGN population. We further investigate how the clustering strength, and hence the DMH mass, varies with $M_{\mathrm{BH}}$, $\lambda_{\mathrm{Edd}}$ and $L_{\mathrm{X}}$. We emphasize that all clustering measurements and bias estimates presented below are derived exclusively from the large-scale ($r_p \gtrsim 2$--$3\,h^{-1}\,\mathrm{Mpc}$) two-halo regime, where the effects of the one-halo term are negligible and linear biasing provides a valid description of the clustering signal. Given the limited AGN statistics after multivariate matching, our binning strategy is intentionally conservative and optimized for the stability of the inferred large-scale bias rather than for the visual smoothness of $w_p(r_p)$.

\begin{figure}
\centering
  \includegraphics[width=0.8\columnwidth, height=5.5cm]{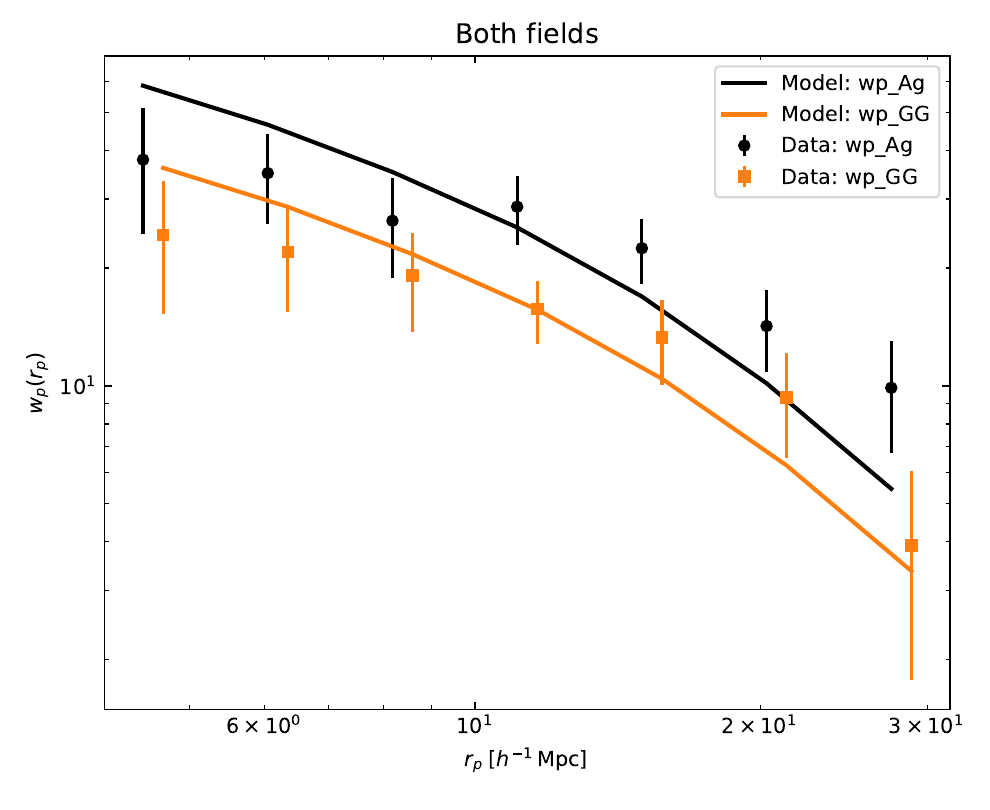}  
  \includegraphics[width=0.8\columnwidth, height=5.5cm]{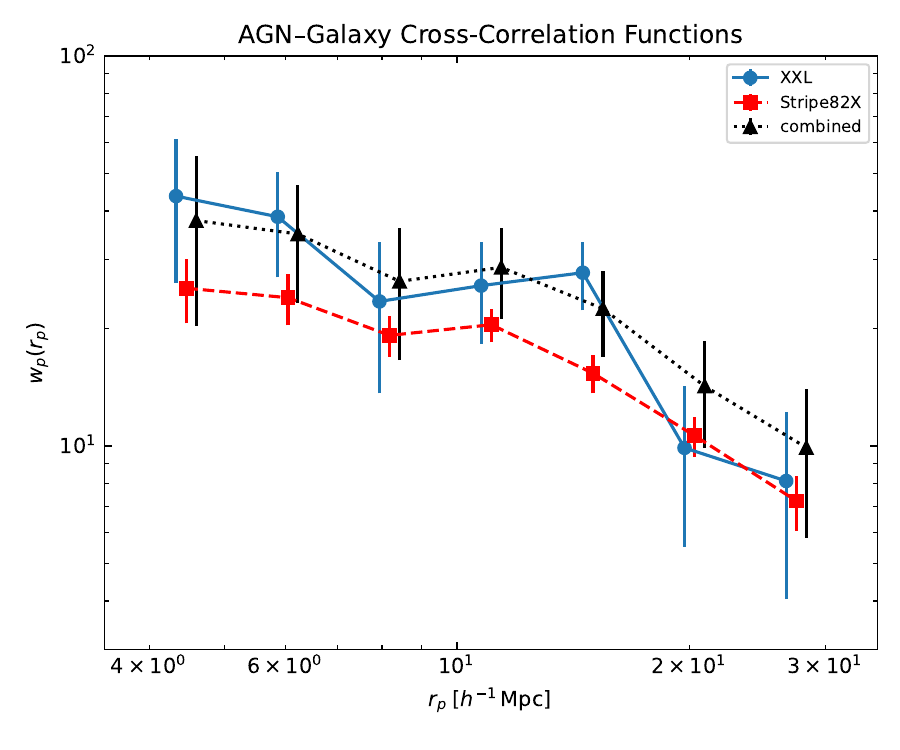}  
   \includegraphics[width=0.8\columnwidth, height=5.5cm]{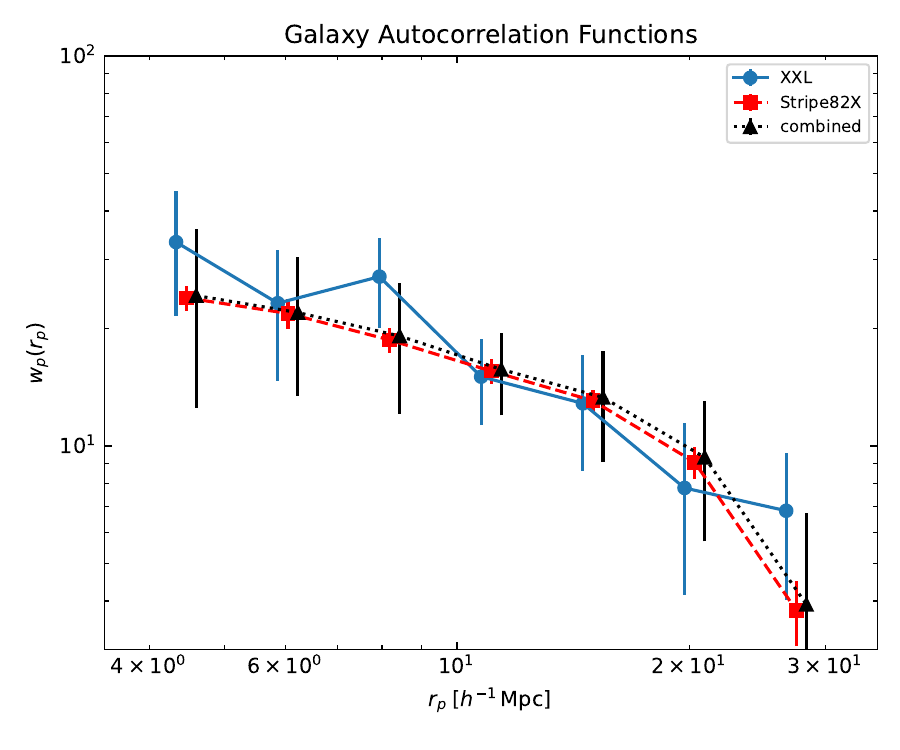}  
  \caption{Correlation function measurements: Top panel: AGN–galaxy cross-correlation function (black points) and galaxy autocorrelation function (orange points) for the combined XXL and Stripe 82X samples, shown together with their best-fitting models (solid lines). Middle and bottom panels: same measurements performed separately for each field, compared with the combined results. The corresponding DMH mass estimates for each field and for the merged dataset are listed in Table~\ref{table_total}. Data points have been offset in the x-axis for clarity.}
  \label{fig_xcross_total}
\end{figure}

\begin{table*}[ht]
\centering
\small
\setlength{\tabcolsep}{0.9pt} 
\caption{Median properties of the AGN and galaxy samples in the XXL, Stripe 82X, and combined fields.}
\label{table_total}
\begin{tabular}{lccccccccccc} 
\hline\hline
Field & Sample & $N_{\mathrm{src}}$ &  $\langle  z \rangle$ & $\langle \log(M_\star/M_\odot) \rangle$&
$\langle \log(\mathrm{SFR}/M_\odot\,\mathrm{yr^{-1}})\rangle$ &
$\langle\log(\mathrm{sSFR}/\mathrm{Gyr^{-1}})\rangle$ &
$\langle\log(L_{\mathrm{X}}/\mathrm{erg\,s^{-1}})\rangle$ &
$\langle\log(M_{\mathrm{BH}}/M_\odot)\rangle$ &
$\langle\log\lambda_{\mathrm{Edd}}\rangle$ &
$\log(M_{\mathrm{DMH}}/h^{-1}M_\odot)$\\
\hline
XXL        & AGN      & 203 & 0.82 & 10.81 & 1.81 &  0.07 & 43.79 & 8.28 & -1.36 & $13.06 \pm 0.27$ \\
           & Galaxies & 2\,428 & 0.65 & 11.02 & 1.67 & -0.27 & -- & -- & -- & $12.98 \pm 0.07$ \\
\hline
Stripe 82X  & AGN      & 245 & 0.82 & 11.17 & 1.83 & -0.38 & 44.04 & 8.60 & -1.58 & $12.83 \pm 0.12$ \\
           & Galaxies & 19\,141 & 0.68 & 10.89 & 1.51 & -0.29 & -- & -- & -- & $12.68 \pm 0.03$ \\
\hline
Combined   & AGN      & 448 & 0.82 & 11.02 & 1.82 & -0.18 & 43.92 & 8.45 & -1.47 & $13.17 \pm 0.22$ \\
           & Galaxies & 21\,569 & 0.68 & 10.91 & 1.52 & -0.29 & -- & -- & -- & $12.74 \pm 0.16$ \\
\hline
\end{tabular}
\tablefoot{
Columns: field, sample type, number of sources ($N_{\mathrm{src}}$), median values of redshift ($\tilde{z}$),
stellar mass ($M_\star$), SFR, sSFR, X-ray luminosity ($L_{\mathrm{X}}$), black hole mass ($M_{\mathrm{BH}}$), Eddington ratio ($\lambda_{\mathrm{Edd}}$),
and measurements of the effective dark matter halo mass ($M_{\mathrm{DMH}}$).
Dashes indicate quantities not applicable to galaxies.
}
\end{table*}

\subsection{Clustering and dark matter halo mass of the full X-ray AGN population}
\label{sec:results_total}

We first estimate the DMH masses of the total AGN (i.e. including quiescent hosts) and galaxy populations in the two fields independently, and then combine the pair counts (DD and DR) from both surveys to derive the overall AGN--galaxy cross-correlation function and the galaxy autocorrelation function \citep[e.g.,][]{Mountrichas2013, Georgakakis2014}. This combined measurement provides the characteristic DMH mass of the X-ray AGN population and of the galaxy tracer sample across both fields.

Figure~\ref{fig_xcross_total} (top panel) presents the AGN–galaxy cross-correlation function and the galaxy autocorrelation function for the combined XXL and Stripe 82X samples, together with their best-fitting models. The fitting is performed over scales of $4$–$30\,h^{-1}\,\mathrm{Mpc}$; however, we verify that the derived results remain statistically consistent when adopting alternative fitting ranges, such as $2$–$25\,h^{-1}\,\mathrm{Mpc}$. The middle and bottom panels show the AGN--galaxy and galaxy autocorrelation functions, respectively, measured separately in each field, as well as the combined results. The properties of the AGN and galaxy datasets, together with the corresponding DMH mass estimates for each field and for the merged sample, are listed in Table~\ref{table_total}. As already mentioned, the combined cross-correlation function is obtained by summing the DD and DR pair counts from the individual fields, using the same
estimator and the same random-to-data ratio in all cases. In this formulation, $w_{p}^{\mathrm{comb}}(r_{p})$ is a DR-weighted average of the individual field measurements, and small deviations where the combined curve lies slightly above or below the individual ones are fully consistent with the statistical uncertainties.

Our results show that X-ray AGN reside in DMHs with typical masses of $\sim10^{13}\,M_\odot$, consistent with previous clustering studies over a wide range of redshifts up to $z\sim3$ \citep[e.g.][]{Allevato2011, Krumpe2012, Mountrichas2012, Mountrichas2013, Koutoulidis2013, Georgakakis2014, Krumpe2018, Viitanen2019}. \citet{Powell2020} investigated the clustering of X-ray selected AGN in the same fields used here (Stripe 82X and XMM--XXL North) by measuring the AGN autocorrelation function in two X-ray luminosity intervals: a low-luminosity bin with $10^{43} \leq L_{\mathrm{X}}\,(\mathrm{2-10\,keV}) < 10^{44.5}\,\mathrm{erg\,s^{-1}}$ at $z \approx 0.8$, and a high-luminosity bin with $L_{\mathrm{X}}\,(\mathrm{2-10\,keV}) > 10^{44.5}\,\mathrm{erg\,s^{-1}}$ at $z \approx 1.8$. Their results for the low-luminosity subset yielded a halo mass of $\log(M_{\mathrm{DMH}}/h^{-1}M_\odot) = 12.83^{+0.25}_{-0.41}$ (see their Table 2), in excellent agreement with our measurements.

\subsection{The relation between black-hole mass and large-scale environment}
\label{sec_mbh}

\begin{figure}
\centering
  \includegraphics[width=0.95\columnwidth, height=6.5cm]{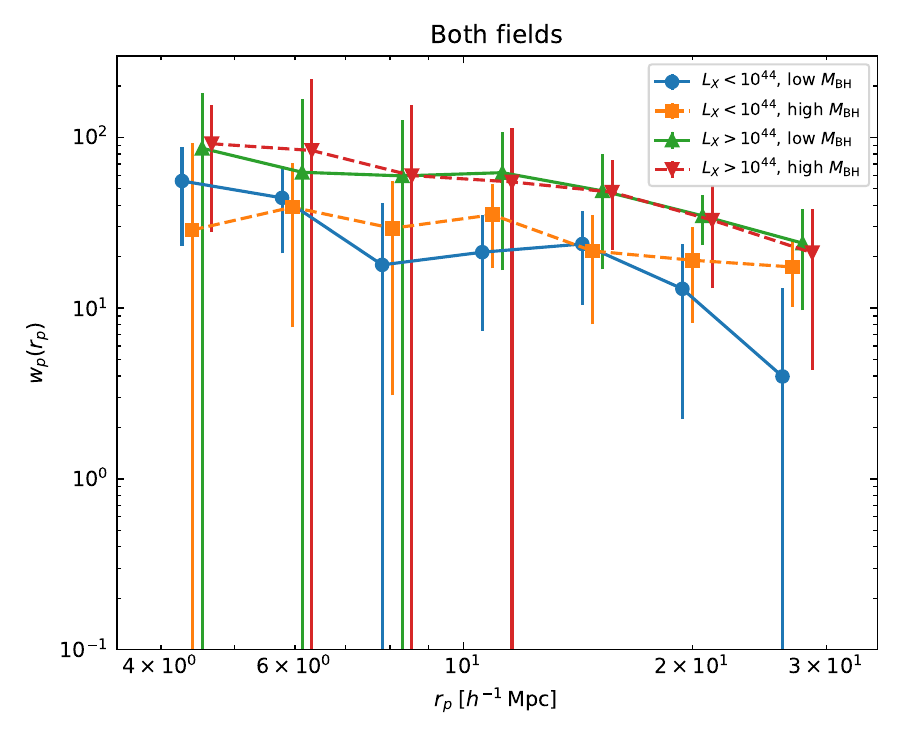}  
  \caption{AGN--galaxy cross-correlation functions as a function of $M_{\mathrm{BH}}$, in two L$_X$ bins, as indicated in the legend. The corresponding DMH mass estimates are listed in Table~\ref{table_mbh}.}
  \label{fig_xcross_mbh}
\end{figure}

\begin{table*}
\centering
\caption{Median properties of the matched AGN subsets used to investigate the dependence of environment on black hole mass.}
\label{table_mbh}
\resizebox{\textwidth}{!}{
\begin{tabular}{lcccccccccc}
\hline\hline
Subset & $N$ & $\langle z \rangle$ &$\langle \log\,{\rm sSFR} \rangle$ & $\langle \log\,{\rm SFR} \rangle$ & $\langle \log\,M_\star \rangle$ & $\langle \log\,L_{\mathrm{X}} \rangle$ & 
$\log(M_{\mathrm{BH}}/M_\odot)$ & $\langle \log\,\lambda_{\mathrm{Edd}} \rangle$ & $b_{\mathrm{AGN}}$ & $\log M_{\mathrm{DMH}}$ \\
\hline
low–$L_{\mathrm{X}}$, low–$M_{\mathrm{BH}}$  & 62 & $0.66$ & $-0.15$ & $1.64$ & $10.89$ & $43.64$ & $8.01$ &  $-1.39$ & $1.75 \pm 0.40$ & $13.10 \pm 0.42$ \\
low–$L_{\mathrm{X}}$, high–$M_{\mathrm{BH}}$ & 62& $0.83$ & $-0.20$ & $1.75$ & $10.96$ & $43.80$ & $8.48$  & $-1.52$ & $1.62 \pm 0.50$ & $12.70 \pm 0.65$ \\
high–$L_{\mathrm{X}}$, low–$M_{\mathrm{BH}}$ & 51& $0.89$ & $+0.02$ & $1.98$ & $10.93$ & $44.19$& $8.43$  & $-1.32$ & $2.33 \pm 0.35$ & $13.29 \pm 0.24$ \\
high–$L_{\mathrm{X}}$, high–$M_{\mathrm{BH}}$ & 51& $0.97$ & $+0.03$ & $1.98$ & $11.09$ & $44.24$ & $8.84$ & $-1.53$ & $2.66 \pm 0.20$ & $13.39 \pm 0.11$ \\
all–$L_{\mathrm{X}}$, low–$M_{\mathrm{BH}}$  &123 & $0.71$ & $-0.14$ & $1.75$ & $10.89$ & $43.85$ & $8.17$ &  $-1.43$ & $1.92 \pm 0.22$ & $13.18 \pm 0.20$ \\
all–$L_{\mathrm{X}}$, high–$M_{\mathrm{BH}}$ & 123& $0.93$ & $-0.13$ & $1.94$ & $11.13$ & $44.11$ & $8.62$  & $-1.69$ & $2.02 \pm 0.30$ & $13.03 \pm 0.29$ \\
\hline
\end{tabular}}
\tablefoot{
Median values of specific star formation rate (sSFR), star formation rate (SFR), 
stellar mass ($M_\star$), X-ray luminosity ($L_{\mathrm{X}}$), redshift ($z$), 
black hole mass ($M_{\mathrm{BH}}$), Eddington ratio ($\lambda_{\mathrm{Edd}}$), 
the AGN linear bias parameter ($b_{\mathrm{AGN}}$), and the inferred dark matter 
halo mass ($M_{\mathrm{DMH}}$) for the matched AGN subsets. 
The bias parameter is derived from the AGN--galaxy cross-correlation analysis, 
and halo masses are obtained by converting $b_{\mathrm{AGN}}$ to 
$M_{\mathrm{DMH}}$ using the Sheth--Tormen formalism assuming $\sigma_8=0.811$. 
For the four subsamples split simultaneously in luminosity and black hole mass, 
``low–high'' divisions correspond to $\log L_{\mathrm{X}} \le 44$ and 
$\log M_{\mathrm{BH}} \le 8.3$ (low) versus $\log L_{\mathrm{X}} > 44$ and 
$\log M_{\mathrm{BH}} > 8.6$ (high). 
For the ``all–$L_{\mathrm{X}}$'' rows, the samples are divided using a single 
black hole mass threshold at $\log(M_{\mathrm{BH}}/M_\odot)=8.5$.}
\end{table*}

In this section, we examine how the typical DMH mass of X-ray selected AGN depends on their $M_{\mathrm{BH}}$. For this purpose, we use the datasets presented in Table~\ref{table_numbers}, i.e., excluding quiescent systems from the AGN samples (see Sect.~\ref{sec_selection_criteria}), and apply the multivariate nearest-neighbour matching algorithm described in Sect.~\ref{sec_matching}.

Ideally, one would like to divide the AGN into low- and high-$M_{\mathrm{BH}}$ subsets while leaving a buffer region between the two to account for the typical uncertainties in $M_{\mathrm{BH}}$, which are on the order of 0.3–0.5\,dex. However, the available sample size does not allow for such fine subdivision. Instead, we use the median $M_{\mathrm{BH}}$ values of the combined X-ray datasets in the low/moderate- and high-luminosity bins, as defined by the cut at $\log(L_{\mathrm{X}}/\mathrm{erg\,s^{-1}})=44$ (see Sect.~\ref{sec_matching}). Specifically, we adopt thresholds of $\log(M_{\mathrm{BH}}/M_\odot)=8.3$ and 8.6 for the low/moderate- and high-$L_{\mathrm{X}}$ AGN, respectively.

When running the matching algorithm, we use sSFR, SFR, and $\lambda_{\mathrm{Edd}}$ as covariates. 
Including additional parameters such as $M_\star$ or $L_X$ would substantially reduce the number of matched sources and compromise the statistical robustness of the clustering measurements. 
Nevertheless, matching on sSFR, SFR, and $\lambda_{\mathrm{Edd}}$ results in AGN subsets with highly similar distributions in $M_\star$, redshift, and $L_X$, as verified a posteriori in Table~\ref{table_mbh}.

The AGN–galaxy cross-correlation functions for the four subsets are shown in 
Fig.~\ref{fig_xcross_mbh}, while the corresponding bias measurements and DMH 
mass estimates are listed in Table~\ref{table_mbh}. In addition to the subsets 
split simultaneously in $L_{\mathrm{X}}$ and $M_{\mathrm{BH}}$, the table also 
includes results for samples divided only by black hole mass across the full 
luminosity range (``all–$L_{\mathrm{X}}$'' rows), providing a complementary 
comparison that maximizes the available statistics.

Due to the small number of sources available in each subset, the fitting is 
performed over scales of $4$–$30\,h^{-1}\,\mathrm{Mpc}$. Based on our 
measurements, the large-scale environment of X-ray AGN shows no significant 
dependence on $M_{\mathrm{BH}}$, for either the low/moderate or the high 
$L_{\mathrm{X}}$ subsamples. We note that individual $w_p(r_p)$ measurements 
exhibit modest bin-to-bin fluctuations, which are expected given the limited 
AGN statistics and the correlated jackknife uncertainties. The inferred bias 
and halo masses are obtained from fits over a broad range of scales and are 
robust to reasonable variations in the fitting range.

\begin{table*}
\centering
\caption{Median properties of the matched AGN subsets used to investigate the dependence of environment on $\lambda_{\mathrm{Edd}}$.}
\label{table_nedd}
\resizebox{\textwidth}{!}{
\begin{tabular}{lcccccccccc}
\hline\hline
Subset & $N$ & $\langle z \rangle$ & $\langle \log\,{\rm sSFR} \rangle$ & $\langle \log\,{\rm SFR} \rangle$ & $\langle \log\,M_\star \rangle$ & $\langle \log\,L_{\mathrm{X}} \rangle$ & 
$\langle \log\,M_{\mathrm{BH}} \rangle$ & $\langle \log\,\lambda_{\mathrm{Edd}} \rangle$ & $b_{\mathrm{AGN}}$ & $\log M_{\mathrm{DMH}}$ \\
\hline
low–$L_{\mathrm{X}}$, low–$\lambda_{\mathrm{Edd}}$   & 66 & 0.70 & $-0.19$ & $1.70$ & $10.93$ & $43.80$ & $8.39$ &  $-1.71$ & $1.63 \pm 0.30$ & $12.94\pm0.38$ \\
low–$L_{\mathrm{X}}$, high–$\lambda_{\mathrm{Edd}}$  & 66 & 0.83 & $-0.17$ & $1.80$ & $10.93$ & $43.81$ & $8.27$ &  $-1.29$ & $2.09 \pm 0.27$ & $13.16\pm0.23$ \\
high–$L_{\mathrm{X}}$, low–$\lambda_{\mathrm{Edd}}$  & 53 & 0.81 & $+0.09$ & $1.94$ & $11.01$ & $44.27$ & $8.71$ &  $-1.67$ & $2.21 \pm 0.36$ & $13.32\pm0.28$ \\
high–$L_{\mathrm{X}}$, high–$\lambda_{\mathrm{Edd}}$ & 53 & 0.91 & $+0.03$ & $2.00$ & $10.90$ & $44.32$ & $8.51$ &  $-1.30$ & $3.14 \pm 0.84$ & $13.72\pm0.36$ \\
all–$L_{\mathrm{X}}$, low–$\lambda_{\mathrm{Edd}}$   & 146 & 0.74 & $-0.16$ & $1.73$ & $10.93$ & $43.96$ & $8.19$ &  $-1.54$ & $1.91 \pm 0.26$ & $13.14\pm0.23$ \\
all–$L_{\mathrm{X}}$, high–$\lambda_{\mathrm{Edd}}$  & 146 & 0.91 & $-0.11$ & $1.99$ & $11.14$ & $44.16$ & $7.92$ &  $-0.71$ & $2.52 \pm 0.35$ & $13.41\pm0.20$ \\
\hline
\end{tabular}}
\tablefoot{
Median values of specific star formation rate (sSFR), star formation rate (SFR), 
stellar mass ($M_\star$), X-ray luminosity ($L_{\mathrm{X}}$), redshift ($z$), 
black hole mass ($M_{\mathrm{BH}}$), Eddington ratio ($\lambda_{\mathrm{Edd}}$), the AGN linear bias parameter ($b_{\mathrm{AGN}}$),
and the inferred dark matter halo mass ($M_{\mathrm{DMH}}$) for the matched AGN 
subsets. ``Low–high'' divisions correspond to $\log L_{\mathrm{X}} \le 44$ and 
$\log \lambda_{\mathrm{Edd}} \le -1.50$ (low) versus $\log L_{\mathrm{X}} > 44$ 
and $\log \lambda_{\mathrm{Edd}} > -1.50$ (high). The same Eddington-ratio 
threshold is used for the ``all–$L_{\mathrm{X}}$'' subsamples. The last column 
lists the DMH masses derived from the AGN--galaxy cross-correlation analysis.
}
\end{table*}

\citet{Krumpe2015} used RASS/SDSS-detected AGN at redshifts $z=0.2$–$0.3$ and reported a weak dependence of clustering strength on $M_{\mathrm{BH}}$. It is important to note, however, several key differences between their study and ours. Their X-ray sample includes sources with systematically lower $M_{\mathrm{BH}}$ (see their Table~3). More crucially, \citet{Krumpe2015} did not account for differences in host-galaxy properties, which can influence the inferred DMH masses \citep[e.g.,][]{Mountrichas2019, Allevato2019}.

\citet{Shen2009} analysed optical quasars from the SDSS Data Release~5 (DR5) and found no significant difference in the clustering strength when splitting their sample into high- and low-$M_{\mathrm{BH}}$ subsamples. However, when they compared the most massive 10\% of black holes to the rest of the population, the quasars hosting the largest $M_{\mathrm{BH}}$ exhibited stronger clustering. It is worth noting, however, that their dataset consists exclusively of optically selected quasars, in contrast to the X-ray–detected AGN studied here. In addition, their analysis covers a broad redshift range ($0.4 < z < 2.5$) and therefore does not account for potential redshift evolution. Finally, their approach does not control for differences in other black hole (e.g., $\lambda_{\mathrm{Edd}}$, luminosity) or host-galaxy properties, which are likely to vary between the low- and high-$M_{\mathrm{BH}}$ subsamples.

\subsection{The relation between Eddington ratio and large-scale environment}
\label{sec_nedd}

\begin{figure}
\centering
  \includegraphics[width=0.95\columnwidth, height=6.5cm]{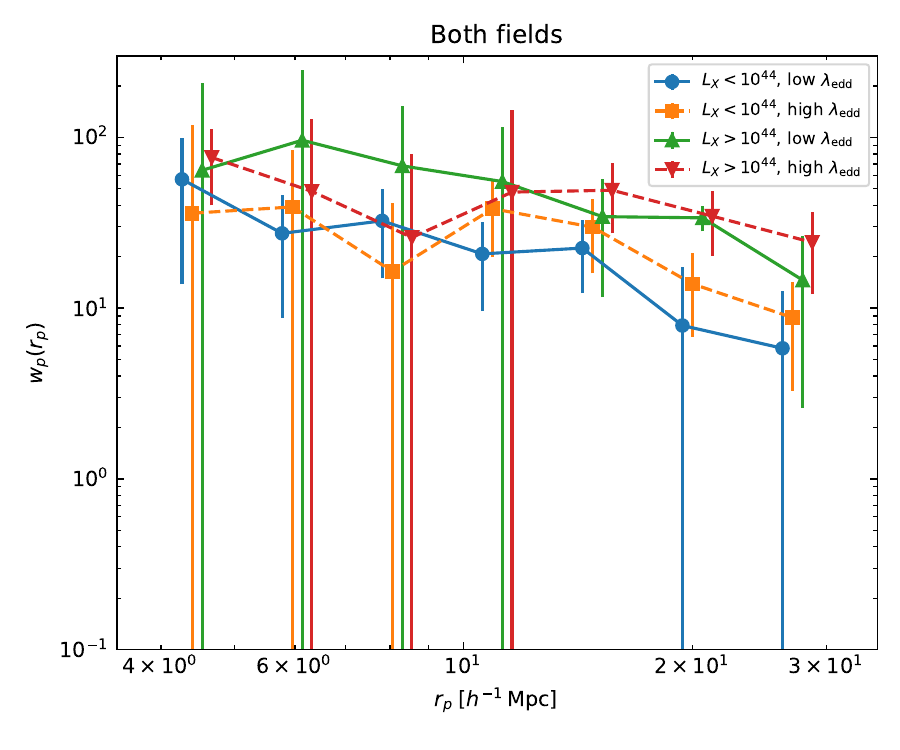}  
  \caption{AGN--galaxy cross-correlation functions as a function of $\lambda_{\mathrm{Edd}}$, in two L$_X$ bins, as indicated in the legend. The corresponding DMH mass estimates are listed in Table~\ref{table_nedd}.}
  \label{fig_xcross_nedd}
\end{figure} 

In this section, we investigate how the DMH mass of X-ray selected AGN depends on their $\lambda_{\mathrm{Edd}}$, using the datasets summarized in Table~\ref{table_numbers} and applying the multivariate nearest-neighbour matching procedure described in Sect.~\ref{sec_matching}. The $\lambda_{\mathrm{Edd}}$ for each AGN  is computed following the standard definition
\begin{equation}
\lambda_{\mathrm{Edd}} = \frac{L_{\mathrm{bol}}}{L_{\mathrm{Edd}}}, \qquad
L_{\mathrm{Edd}} = 1.26 \times 10^{38} \left( \frac{M_{\mathrm{BH}}}{M_\odot} \right)\, \mathrm{erg\,s^{-1}}.
\end{equation}
In the implementation used here, $L_{\mathrm{bol}}$ is derived from the SED fitting.

Similarly to the previous section, we adopt a luminosity threshold at $\log(L_{\mathrm{X}}/\mathrm{erg\,s^{-1}})=44$ to define low/moderate- and high-luminosity AGN. Within each luminosity bin, the median $\log \lambda_{\mathrm{Edd}}$ of the combined XXL and Stripe 82X datasets is $-1.5$, and this value is used to separate AGN into low- and high-$\lambda_{\mathrm{Edd}}$ subsets. The multivariate matching algorithm is applied to the star-forming AGN populations within each luminosity bin, using $\log \mathrm{sSFR}$, $\log \mathrm{SFR}$, and $\log M_{\mathrm{BH}}$ as matching covariates. The resulting matched sample properties are listed in Table~\ref{table_nedd}.

The AGN–galaxy cross-correlation functions for the four matched subsets are shown in Fig.~\ref{fig_xcross_nedd}. The calculated $b_{AGN}$ and the corresponding DMH mass estimates are listed in Table~\ref{table_nedd}. The fits are performed over scales of $4$–$30\,h^{-1}\,\mathrm{Mpc}$. We find no statistically significant dependence of the typical halo mass on $\lambda_{\mathrm{Edd}}$, with any apparent trend being consistent with statistical fluctuations at the $\sim1\sigma$ level. Previous clustering studies reported little or no correlation between halo mass and $\lambda_{\mathrm{Edd}}$ \citep[e.g.][]{Krumpe2015}, or with its empirical proxy, the specific black-hole accretion rate ($L_{\mathrm{X}}/M_\star$; \citealt{Allevato2019, Viitanen2019}).


\begin{table*}
\centering
\caption{Median properties of the matched AGN subsets used to investigate the dependence of environment on L$_X$.}
\label{table_lx}
\resizebox{\textwidth}{!}{
\begin{tabular}{lcccccccccc}
\hline\hline
Subset & $N$ & $\langle z \rangle$ & $\langle \log\,{\rm sSFR} \rangle$ & $\langle \log\,{\rm SFR} \rangle$ & $\langle \log\,M_\star \rangle$ & $\langle \log\,L_{\mathrm{X}} \rangle$ & 
$\langle \log\,M_{\mathrm{BH}} \rangle$ & $\langle \log\,\lambda_{\mathrm{Edd}} \rangle$ & $b_{\mathrm{AGN}}$ & $\log M_{\mathrm{DMH}}$ \\
\hline
low–$L_{\mathrm{X}}$  & 115 & $0.79$ & $-0.14$ & $1.84$ & $10.96$ & $43.82$ & $8.52$ & $-0.91$ & $ 2.04\pm0.39$  & $ 13.21\pm0.31 $ \\
high–$L_{\mathrm{X}}$ & 115 & $0.83$ & $-0.03$ & $1.87$ & $10.97$ & $44.19$ & $8.54$ & $-0.86$ & $ 2.41\pm0.24 $ & $ 13.42\pm0.14 $ \\
\hline
\end{tabular}}
\tablefoot{
Median values of specific star formation rate (sSFR), star formation rate (SFR), stellar mass ($M_\star$), X-ray luminosity ($L_{\mathrm{X}}$), redshift ($z$), black hole mass ($M_{\mathrm{BH}}$), Eddington ratio ($\lambda_{\mathrm{Edd}}$), and measurements of the AGN linear bias parameter ($b_{\mathrm{AGN}}$), and the effective dark matter halo mass ($M_{\mathrm{DMH}}$) for the two matched AGN subsets.  
``Low–high'' divisions correspond to $\log L_{\mathrm{X}} \le 44$  (low) versus $\log L_{\mathrm{X}} > 44$ (high).}
\end{table*}

\subsection{The relation between X-ray luminosity and large-scale environment}
\label{sec_lx}

\begin{figure}
\centering
  \includegraphics[width=0.85\columnwidth, height=6.5cm]{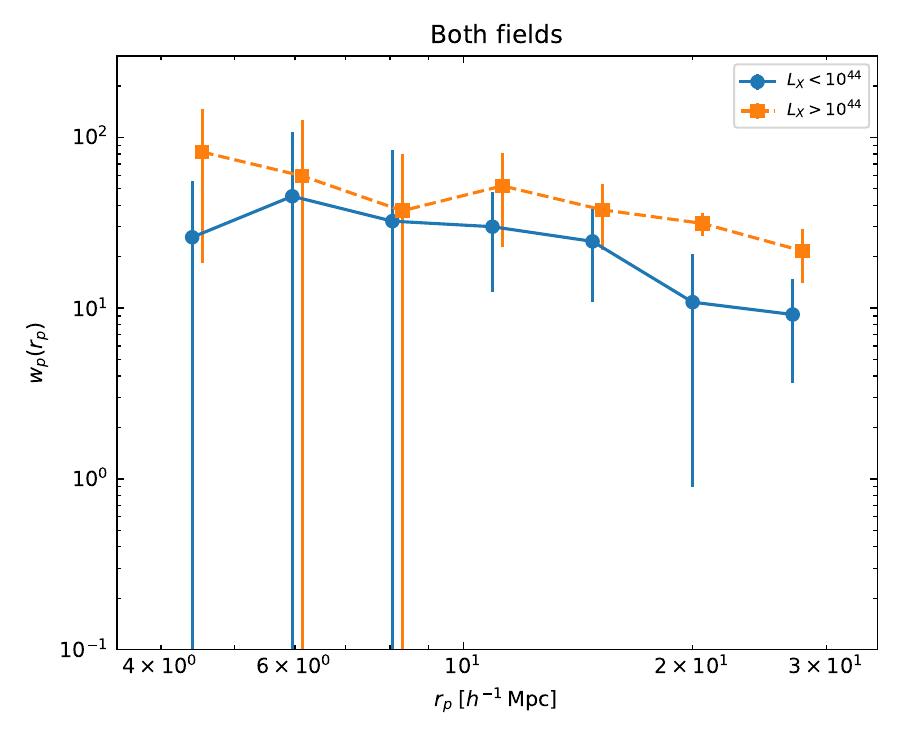}  
  \caption{AGN--galaxy cross-correlation functions as a function of $L_X$. The corresponding DMH mass estimates are listed in Table~\ref{table_lx}.}
  \label{fig_xcross_lx}
\end{figure}

\begin{figure}
\centering
  \includegraphics[width=0.85\columnwidth, height=6.5cm]{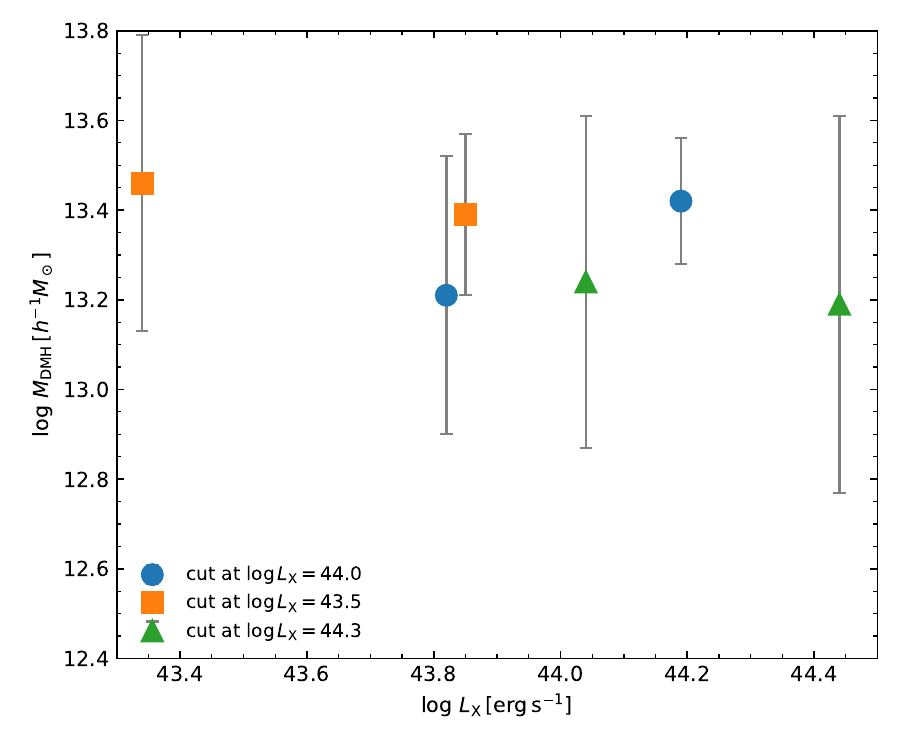}  
  \caption{
Dependence of the characteristic DMH mass on the $L_X$ 
threshold used to divide the AGN sample. Each symbol type corresponds to a different luminosity cut applied to split the Because the sSFR bins are defined by quantiles, the statistical
uncertainties are of similar order across bins; the highest-sSFR bin is
open-ended and therefore represents a broader range of star-formation
activity.
sample into ``low'' and ``high'' $L_{\mathrm{X}}$ subsamples: 
$\log L_{\mathrm{X}}=43.5$ (squares), $44.0$ (circles), and $44.3$ (triangles). 
For each luminosity threshold, two data points are shown, representing the 
low-$L_{\mathrm{X}}$ and high-$L_{\mathrm{X}}$ subsamples defined by that cut. 
The horizontal position of each point corresponds to the median X-ray luminosity 
of the respective subsample. Error bars indicate the $1\sigma$ uncertainties on 
the derived dark matter halo mass.}
  \label{fig_dmhm_lx}
\end{figure}

In this section, we examine how the DMH mass of X-ray selected AGN depends on their $L_{\mathrm{X}}$. The \citep[e.g.][]{Shankar2013, Aird2012, Allevato2011, Georgakakis2019}.multivariate nearest-neighbour matching algorithm (see Sect.~\ref{sec_matching}) is applied to the AGN samples using $\log\mathrm{sSFR}$, $\log\mathrm{SFR}$, $\log\lambda_{\mathrm{Edd}}$, and $\log M_{\mathrm{BH}}$ as covariates. The median properties of the resulting matched subsets are listed in Table~\ref{table_lx}.

The AGN–galaxy cross-correlation functions for the matched samples are shown in Fig.~\ref{fig_xcross_lx}. The calculated $b_{AGN}$ and the corresponding DMH mass estimates are presented in Table~\ref{table_lx}. Our measurements show a hint that more luminous AGN tend to reside in more massive DMHs compared to their less luminous counterparts, however the statistical significance of this trend is statistically insignificant. 

Previous studies have reported mixed results regarding the dependence of AGN clustering on luminosity. Several works have found a positive correlation, with more luminous AGN inhabiting more massive halos \citep[e.g.][]{Krumpe2010a, Krumpe2012, Koutoulidis2013}, while others reported no significant dependence on large scales, but differences emerging at small scales \citep[][]{Krumpe2018}. As discussed earlier, the limited size of our current datasets does not allow us to probe the small-scale regime of the correlation function. Interestingly, an opposite (negative) luminosity dependence has also been observed, with higher-luminosity AGN found in less massive halos \citep[][]{Mountrichas2016}. That discrepancy was attributed to differences in the luminosity ranges explored across various studies, an interpretation supported by semi-analytic model predictions \citep{Fanidakis2013a,Fanidakis2013b}; see also Fig.~7 of \citet{Mountrichas2016}. Finally, more recent studies have claimed that the clustering differences between different AGN selection techniques are dominated by selection biases, and not due to a dependence on AGN luminosity \citep[][]{Powell2020}.

To further assess whether the inferred halo masses depend on AGN luminosity, we recomputed the clustering measurements after applying additional luminosity thresholds at $\log L_{\mathrm{X}} = 43.5$ and $\log L_{\mathrm{X}} = 44.3$. A more stringent cut at $\log L_{\mathrm{X}}=44.5$ was not pursued, as the multivariate matching leaves too few AGN for a reliable measurement.  
The results for all luminosity-defined subsets are presented in 
Fig.~\ref{fig_dmhm_lx}.

Measurement uncertainties in $\log L_{\mathrm{X}}$, together with the multivariate matching procedure, can introduce partial cross-talk between the luminosity subsamples at the population level. As a result, formal correlation coefficients or regression analyses between halo mass and $L_{\mathrm{X}}$ would not provide meaningful constraints, since the effective number of independent degrees of freedom is reduced. In this context, the appropriate comparison is the qualitative and quantitative consistency of the halo-mass estimates across different luminosity cuts, rather than attempting to fit a continuous trend.

Across all luminosity thresholds explored, the inferred halo masses remain 
consistent within their uncertainties, with no systematic increase or decrease 
toward higher $L_{\mathrm{X}}$. This stability across overlapping luminosity selections indicates that, within our sample, the large-scale environments of X-ray AGN show no statistically significant correlation with $L_{\mathrm{X}}$.

\section{Discussion}
\label{sec_discussion}

The analysis presented in this work is performed within a well-defined and physically motivated region of parameter space. By focusing on X-ray selected AGN at fixed redshift ($0.5<z<1.2$) and controlling for host-galaxy properties, we probe the typical regime of massive, star-forming AGN hosts rather than the full galaxy population. This approach minimizes evolutionary mixing and isolates secondary dependencies beyond those set by the host galaxy.

Based on our results, we find no statistically significant correlation between DMH mass and
$M_{\mathrm{BH}}$ across the range
$7.5 \lesssim \log(M_{\mathrm{BH}}/M_\odot) \lesssim 10$.
Given the limited dynamic range in $M_{\mathrm{BH}}$ and the modest AGN statistics after multivariate matching, this result should be interpreted as the absence of a detectable trend within the uncertainties of the present dataset, rather than as evidence for the intrinsic absence of a coupling between $M_{\mathrm{BH}}$ and halo mass.

The relatively limited dynamic range in $M_{\mathrm{BH}}$ primarily reflects the de\citep[e.g.][]{Shankar2013, Aird2012, Allevato2011, Georgakakis2019}.liberate choice to focus on a narrow redshift interval. In this regime, the analysis is optimized to test for the presence of strong or systematic differences in halo mass at fixed host-galaxy properties, rather than to constrain weak or continuous dependencies that would require substantially larger samples and broader parameter coverage. In this context, the absence of statistically significant variations is itself informative, indicating that any residual dependence of halo mass on black-hole or accretion properties must be sub-dominant relative to the effects of the host galaxy.

Several recent studies have suggested that halo mass may play a more fundamental role in regulating black-hole growth than host-galaxy properties alone. For example, clustering and residual analyses indicate that $M_{\mathrm{BH}}$ may correlate with halo mass at fixed $M_\star$ or velocity dispersion \citep[e.g.][]{Powell2020, Shankar2025},
while semi-analytic models predict a close connection between black-hole growth and halo
assembly history
\citep[e.g.][]{Shankar2010, Shankar2013, Shankar2016, Shankar2017, Menci2023}.
Our analysis does not contradict these findings; rather, it indicates that any such dependence
is likely weak or second-order on the large scales probed here and difficult to detect with
current X-ray AGN samples.

Hydrodynamical simulations such as \textsc{IllustrisTNG} and \textsc{EAGLE} suggest that
the mass of the central black hole correlates most strongly with the depth of the central
gravitational potential of the host galaxy (e.g. its binding energy or velocity dispersion),
while the halo mass influences black-hole growth more indirectly by regulating gas supply,
AGN duty cycle, and long-term evolutionary pathways
\citep[e.g.][]{Booth2010, Dubois2012, Bower2017, Weinberger2018, Terrazas2020, Habouzit2022, Menci2023}.

In this framework, halo mass primarily sets the boundary conditions for black-hole growth
through its impact on galaxy assembly and gas accretion histories, whereas the instantaneous
accretion rate and observed AGN properties are governed predominantly by baryonic processes
operating on galactic and sub-galactic scales.
Our results are consistent with this picture, in which halo mass modulates black-hole growth
indirectly via host-galaxy evolution rather than through a tight, one-to-one correspondence
between halo mass and black-hole mass.
\citep[e.g.][]{Shankar2013, Aird2012, Allevato2011, Georgakakis2019}.

We also note that the low AGN duty cycle expected at $z \lesssim 1$ implies that only a
small fraction of massive halos host an active nucleus at any given epoch \citep[e.g.][]{Allevato2011, Aird2012,Shankar2013, Georgakakis2019}. As a result, even if halo mass plays a fundamental role in black-hole growth over cosmic time, large-scale clustering measurements of active systems are expected to show only weak dependencies once host-galaxy properties are controlled.

$\lambda_{\mathrm{Edd}}$ tracing the instantaneous efficiency of accretion, shows only a weak and statistically insignificant dependence on halo mass. This suggests that the short-term fueling efficiency of AGN is largely independent of the surrounding DMH potential. Large-scale environment may regulate whether a galaxy has gas available for accretion in a statistical sense, but it does not control how efficiently that gas is converted into radiation once it reaches the central regions.  
Our results are consistent with previous studies showing similar halo masses across wide $\lambda_{\mathrm{Edd}}$ ranges \citep[e.g.][]{Krumpe2015, Viitanen2019}, and together they indicate that AGN variability is dominated by stochastic or secular processes within galaxies rather than by their position in the cosmic web. The weak, marginal trend toward higher halo masses at larger $\lambda_{\mathrm{Edd}}$, if real, could reflect the tendency of gas-rich group environments to promote more efficient inflows or longer duty cycles, but this effect is at best secondary.

The comparison between halo mass and $L_{\mathrm{X}}$ reveals no statistically significant correlation. This agrees with previous clustering analyses \citep[e.g.][]{Krumpe2018, Powell2020}, which similarly found that AGN with different X-ray powers occupy halos of comparable mass. This behaviour suggests that once an AGN phase is triggered, its radiative output is primarily governed by internal processes, such as stochastic fueling, accretion-rate variability, or self-regulating feedback, rather than by the large-scale environment. In the luminosity regime probed here ($43 \lesssim \log(L_{\mathrm{X}}/\mathrm{erg\,s^{-1}}) \lesssim 45$), most systems likely operate in a moderate-accretion or “maintenance-mode” regime, where black holes are sustained by intermittent inflows of cold gas but limited by feedback that stabilizes the host’s gas reservoir. As a result, fluctuations in $L_{\mathrm{X}}$ are expected to trace short-term variability or duty-cycle modulation rather than systematic differences in halo potential.

Recent clustering measurements of X-ray AGN in the COSMOS field by 
\citet{Ikeda2025} probe a similar redshift range ($0.6<z<1.4$) but at 
significantly lower X-ray luminosities ($\langle \log L_X \rangle \sim 42.7$). 
They report a characteristic DMH mass of 
$\log M_{\rm DMH} \sim 11.8\,h^{-1}M_\odot$ (AGN bias of $\sim 1.2$), which is substantially lower than 
the values derived in our analysis ($\log M_{\rm DMH} \sim 13.2\,h^{-1}M_\odot$ and AGN bias of $\sim 2.1$). Although the comparison is not fully 
controlled because their sample includes both type~1 and type~2 AGN and does 
not match host-galaxy properties, the difference is consistent with the 
possibility of increasing halo mass with AGN luminosity when considering a 
broader $L_X$ range. We note, however, that halo masses inferred from clustering 
depend on the adopted bias--halo mass relation. \citet{Ikeda2025} use the 
prescription of \citet{Tinker2005}, whereas in this work we convert bias to halo 
mass using the Sheth--Tormen formalism. Consequently, part of the difference in 
the reported halo masses may reflect the different bias--halo mass conversions, 
in addition to differences in the underlying AGN selection and luminosity range.

An additional parameter that may influence measured halo masses is AGN obscuration. Earlier works suggested that obscured AGN may reside in denser environments than unobscured ones \citep[e.g.][]{Allevato2011}, although this has not been universally confirmed \citep[e.g.][]{Viitanen2023}. Our sample consists exclusively of Type~1 AGN, and although optical classification does not perfectly correspond to X-ray absorption, both the XXL and Stripe~82X surveys are relatively shallow, wide-area fields that contain few heavily obscured sources \citep{Liu2016, Peca2023}. Specifically, only 17 ($\sim8\%$) and 12 ($\sim6\%$) of the sources in the XXL and Stripe~82X samples, respectively, have $N_{\mathrm{H}} > 10^{22}\,\mathrm{cm}^{-2}$. Consequently, the limited presence of obscured AGN likely minimizes any environmental bias that could otherwise arise from obscuration-dependent selection effects.

It is important to emphasize that our conclusions are restricted to the large-scale environment probed by the clustering analysis ($\gtrsim1$\,Mpc). While we find no evidence that large-scale environment plays a dominant role in regulating AGN activity once host-galaxy properties are controlled, this does not exclude the importance of environmental processes operating on smaller scales. Physical mechanisms such as galaxy interactions, mergers, or cluster-specific processes occur on $\lesssim$\,Mpc scales, which are not accessible with the present dataset. Numerous studies have demonstrated that such small-scale environmental effects can be relevant in dense environments, particularly in galaxy groups and clusters. Our results therefore indicate that, on large scales, AGN activity is primarily linked to host-galaxy properties, while environmental effects on smaller scales may still contribute to AGN triggering in specific regimes.

Our results integrate naturally with the environmental trends identified in \citet{Mountrichas2019}, where clustering amplitude was found to (mildly) increase with $M_\star$ but decrease with both SFR and sSFR. That earlier study demonstrated that dense environments preferentially host massive yet less star-forming galaxies, implying that large-scale structure regulates the availability of cold gas and the efficiency of star formation. The present findings extend this picture by showing that, once galaxies enter an active accretion phase, the large-scale environment no longer determines the level of AGN activity. Instead, the efficiency and luminosity of accretion appear to be internally controlled, while the halo environment mainly influences how often and for how long such phases can occur. Taken together, the two studies suggest an evolutionary continuity: during the star-forming phase, the halo influences gas supply and the likelihood of triggering AGN activity, whereas once accretion begins, the subsequent growth of the black hole and its radiative output are predominantly driven by internal processes within the host.

It is worth emphasizing that, once host-galaxy properties such as $M_\star$ are controlled, the range of halo masses probed by AGN samples is expected to be relatively narrow in a statistical sense. This reflects the underlying monotonic relation between $M_\star$ and halo mass established for the general galaxy population, albeit with substantial intrinsic scatter. In this framework, AGN clustering measurements are not designed to identify triggering mechanisms, but rather to test whether different AGN selections preferentially populate distinct regions of the underlying halo-mass distribution beyond what is already set by their hosts. The absence of significant clustering differences at fixed host properties therefore indicates that $M_{\mathrm{BH}}$, $\lambda_{\mathrm{Edd}}$ and radiative output do not introduce an additional dependence on halo mass beyond that already encoded by the host galaxy.

Overall, our results point toward a coherent physical picture in which the typical dark-matter halo mass of X-ray AGN is of order $\sim10^{13}\,h^{-1}M_\odot$, with no statistically significant variation as a function of $M_{\mathrm{BH}}$, $\lambda_{\mathrm{Edd}}$, or $L_{\mathrm{X}}$ within the uncertainties and parameter ranges probed here. 
While we do not detect a clear $M_{\mathrm{halo}}$–$M_{\mathrm{BH}}$ trend, this should not be interpreted as evidence that such a relation is intrinsically absent. Given the relatively narrow dynamic range in $M_{\mathrm{BH}}$ explored by our sample, combined with observational uncertainties and intrinsic scatter, such a correlation could plausibly remain hidden in large-scale clustering measurements. 

Similarly, the absence of a statistically significant dependence of DMH mass on $\lambda_{\mathrm{Edd}}$ suggests that the instantaneous accretion state is not tightly coupled to halo mass on the scales probed here. 
Instead, large-scale environment appears to regulate AGN activity primarily through its impact on gas supply, star-formation history, and duty cycle, rather than directly controlling short-term accretion efficiency. 
When considered alongside previous evidence for a strong inverse relation between star-formation activity and clustering strength, our findings support a self-regulated co-evolution framework in which halo mass sets broad boundary conditions for galaxy evolution, while black-hole growth and accretion variability are governed predominantly by baryonic processes operating on galactic and sub-galactic scales.

\section{Summary}
\label{sec_summary}

We have investigated the relation between the large-scale environment of X-ray selected AGN and fundamental black-hole and accretion properties, combining the wide-area XXL and Stripe~82X surveys. The joint datasets includes more than $400$ AGN with reliable spectroscopic redshifts ($0.5 \lesssim z \lesssim 1.2$) and $M_{\mathrm{BH}}$ measurements and about $20\,000$ non-AGN galaxies. Physical parameters for all sources have been derived through consistent SED fitting using the \textsc{CIGALE} code.
 
To quantify the large--scale environment of AGN, we measured the AGN–galaxy cross-correlation and galaxy autocorrelation functions and inferred the corresponding DMH masses via standard bias conversion.  
The analysis was performed separately in bins of X-ray luminosity, black-hole mass, and Eddington ratio.  
To ensure that any detected trends reflect genuine physical dependencies rather than host or selection biases, we implemented a multivariate nearest-neighbour matching technique.  
This method pairs AGN subsets while controlling for key covariates such as $M_\star$, SFR, and sSFR (and, when relevant, $M_{\mathrm{BH}}$, $\lambda_{\mathrm{Edd}}$ or $L_X$), thereby isolating the role of the targeted parameter under fixed host conditions.  

\medskip
\noindent
The main results can be summarized as follows:
\begin{itemize}
\item The typical DMH mass of X-ray AGN is $\sim10^{13},h^{-1}M_\odot$, consistent across both \textit{XXL} and \textit{Stripe82X}, indicating that moderate-luminosity AGN inhabit group-sized DMHs.
\item No statistically significant correlation is found between DMH mass and $M_{\mathrm{BH}}$ over the range $7.5 \lesssim \log(M_{\mathrm{BH}}/M_\odot) \lesssim 10$, suggesting that black-hole growth is largely decoupled from the large-scale halo environment once galaxies reach the massive, star-forming regime.
\item The $\lambda_{\mathrm{Edd}}$ shows no measurable correlation with halo mass, indicating that accretion efficiency is largely independent of large-scale environment.
\item The relation between halo mass and X-ray luminosity is statistically flat, indicating that instantaneous AGN power is not strongly correlated with the mass of the surrounding DMH.
\end{itemize}

\medskip
Taken together, these results support a coherent evolutionary picture in which the environment primarily sets the initial gas supply and the likelihood of triggering nuclear activity, while the subsequent growth and variability of the black hole are governed by processes internal to the host.  
Once an AGN phase is triggered, feedback and short-term accretion variability dominate over large-scale environmental effects, yielding similar halo masses across the AGN population.  
This self-regulated framework links galaxy quenching and black-hole fueling as consecutive manifestations of the same baryon-cycle processes operating across scales.  

Future wide-field surveys will enable decisive progress in this direction. 
In particular, the eROSITA all-sky survey \citep{Merloni2024} 
is providing the largest homogeneous samples of X-ray selected AGN to date, 
extending clustering studies over wide luminosity and redshift ranges. 
On the spectroscopic side, large-scale programs such as the SDSS-V Black Hole 
Mapper \citep{Anderson2023, Aydar2025}, the Dark Energy Spectroscopic 
Instrument (DESI) survey \citep{DESI2016}, the Subaru Prime Focus Spectrograph 
(PFS) \citep[e.g.][]{Takada2014}, and the forthcoming 4MOST surveys 
\citep[e.g.][]{Jong2012} will deliver extensive redshift measurements and 
broad-line AGN classifications over thousands of square degrees.

The combination of these X-ray and spectroscopic datasets will provide the 
statistical power and uniformity needed to probe the dependence of AGN 
clustering on black-hole mass, accretion rate, and luminosity with 
unprecedented precision. Such samples will allow environmental trends to be 
examined as a continuous function of accretion state and redshift, offering a 
definitive test of the role of large-scale structure in black-hole–galaxy 
co-evolution.

\begin{acknowledgements}
GM acknowledges funding from grant PID2021-122955OB-C41 funded by MCIN/AEI/10.13039/501100011033 and by “ERDF/EU”. This work was partially supported by the European Union's Horizon 2020 Research and Innovation program under the Maria Sklodowska-Curie grant agreement (No. 754510). This publication is part of the R\&D\&I project PID2024-155779OB-C31, funded by MICIU/AEI/10.13039/501100011033 and co-funded by FEDER, EU. We also acknowledge partial support from the European Union’s Horizon 2020 research and innovation pro-
gramme under the Marie Skłodowska-Curie grant agreement No 860744 (Bid4BESt; grant coordinator F. Shankar

\end{acknowledgements}

\bibliography{mybib}
\bibliographystyle{aa}

\appendix

\section{Validation of the angular selection}
\label{sec_appendix}

As an empirical validation of the assumptions underlying the AGN–galaxy
cross-correlation approach, we test whether the angular clustering of the AGN
sample is consistent with that expected for random subsamples of the galaxy
catalogue once the survey mask is taken into account. We compute the AGN
angular two-point correlation function, $w_{\rm AGN}(\theta)$, using the galaxy
random catalogue to describe the angular selection function. We then draw
$N$ random subsamples from the galaxy catalogue, each containing the same
number of objects as the AGN sample and matched to its redshift distribution
in coarse bins, and measure $w(\theta)$ for each subsample using the same
random catalogue.

\begin{figure}[h]
\centering
\includegraphics[width=0.45\textwidth, height=7.5cm]{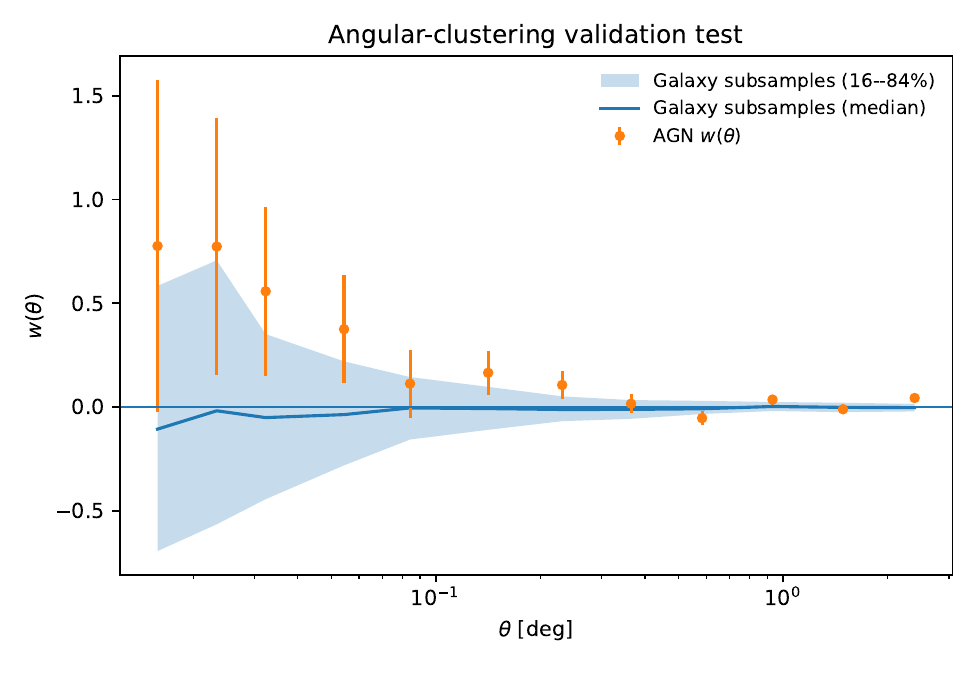}
\caption{Angular correlation functions. Comparison between the AGN angular correlation function, $w_{\rm AGN}(\theta)$,
and the distribution of $w(\theta)$ measured from random galaxy subsamples of
equal size and matched redshift distribution. The shaded region shows the
16--84\,per\,cent range and the solid line the median of the galaxy subsamples.
The AGN measurements are consistent with this distribution on the angular scales relevant for the clustering analysis.}
\label{fig_appendix_wtheta}
\end{figure}

Figure \ref{fig_appendix_wtheta} compares $w_{\rm AGN}(\theta)$ to the distribution of
$w(\theta)$ obtained from the galaxy subsamples. The AGN measurements lie
within the 16--84\,per\,cent range of the galaxy-subsample distribution over
the angular scales corresponding to the two-halo regime probed in this work.
This indicates that, once the survey mask is accounted for, the angular
selection of the AGN sample does not introduce a detectable large-scale
systematic in the inferred clustering amplitude. We stress that this test
does not demonstrate that the AGN and galaxy selection functions are
identical; rather, it provides empirical support for the validity of using
the galaxy random catalogue in the AGN–galaxy cross-correlation analysis on
the large scales relevant for this study.

\end{document}